\documentclass[12pt,preprint]{aastex}
\usepackage{fullpage}
\usepackage[utf8x]{inputenc}
\usepackage{epsf,amsfonts,amsmath,amssymb}
\usepackage{graphicx}
\usepackage{times}
\usepackage{natbib}
\usepackage{ulem}
\usepackage{epsfig}
\usepackage{mathrsfs}
\usepackage{color}

\def\G{\Gamma}

\begin{document}
\title{Constraints on the Synchrotron Emission Mechanism in GRBs}
\author{Paz Beniamini\altaffilmark{1a} \& Tsvi Piran\altaffilmark{1b}}
\altaffiltext{1}{Racah Institute for Physics, The Hebrew University, Jerusalem, 91904, Israel}
\email{(a) paz.beniamini@mail.huji.ac.il; (b) tsvi.piran@mail.huji.ac.il}

\begin{abstract}
We reexamine the general synchrotron model for GRBs' prompt emission and determine the
regime in the parameter phase space in which it is viable.
We characterize a typical GRB pulse in terms of its peak energy, peak flux and duration and
use the latest Fermi observations to constrain the high energy part of the spectrum.
We solve for the intrinsic parameters at the emission region and find the possible parameter phase space 
for synchrotron emission.
Our approach is general and it does not depend on a specific energy dissipation mechanism. 
Reasonable synchrotron solutions are found with energy ratios of $10^{-4}<\epsilon_B/\epsilon_e<10$, bulk Lorentz factor values of $300<\Gamma<3000$,
typical electrons' Lorentz factor values of $3\times 10^3<\gamma_e<10^5$ and emission radii of the order
$10^{15}\mbox{cm}<R<10^{17}$cm. Most remarkable among those are the rather large values of the emission radius and the electron's Lorentz factor.
We find that soft (with peak energy less than 100KeV) but luminous (isotropic luminosity of $~1.5 \times 10^{53}$) pulses  are inefficient.
This may explain the lack of strong soft bursts.
In cases when most of the energy is carried out by the kinetic energy of the flow, such as in the internal shocks,
the synchrotron solution requires that only a small fraction of the electrons are accelerated to relativistic velocities by the shocks.
We show that future observations of very high energy photons from GRBs by CTA, could possibly determine all parameters of
the synchrotron model or rule it out altogether.
\end{abstract}


\section{Introduction}
\label{int}
GRBs' non thermal spectrum has lead to the suggestion that both the prompt and the
afterglow are produced by synchrotron emission of relativistic electrons \citep{Katz(1994),Meszaros(1994),Sari(1996),Sari(1998)}.
There is reasonable agreement between the predictions \citep{Sari(1998)} of the synchrotron model and
afterglow observations \citep{Wijers_Galama(1999),Granot(1999),Granot(2002),Panaitescu(2001),Nousek(2006),Zhang(2006)}.
However, the situation concerning the prompt emission is more complicated.

The observations of many
bursts whose lower spectral slope is steeper than the ``synchrotron line of death" \citep{Crider(1997), Preece(1998), Preece(2002)} pose a serious problem to synchrotron emission.
This has motivated considerations of photospheric thermal emission 
\citep{EichlerLevinson(2000),Meszaros(2000),Peer(2006), Rees(2005), Giannios(2006), Thompson(2007), RydePeer(2009)},
in which thermal energy stored in the bulk is radiated in the prompt phase at the Thomson photosphere and the high energy tail is produced by inverse Compton. 
While this model yields a consistent low energy slope it has its own share of problems. Most notably \cite{Vurm(2012)} have recently shown that
unless the outflow is moving very slowly no known mechanism can produce the needed soft photons. Furthermore, it is not clear how the high energy GeV emission can be produced
within an optically thick regime \citep{Vurm(2012a)}. On the other hand,
observations of some bursts, e.g. GRB 080916C in which there is a strong upper limit on the thermal component \citep{Zhang(2009)}, or
GRBs: 100724B, 110721A, 120323A, where a thermal component was possibly detected but with only a small fraction ($5-10\%$) of the total energy in the thermal component
\citep{Guiriec(2011),Guiriec(2012),Axelsson(2012)}
and other bursts whose spectral slopes is consistent with synchrotron suggest that for some bursts 
synchrotron might be a viable model.
In fact, in all cases that involve Poynting flux outflows that are Poynting flux dominated at the emitting region synchrotron 
is inevitable and possibly dominant \citep{Beniamini(2013)}.
As such it is worthwhile to explore the conditions needed to generate the
observed prompt emission via synchrotron\footnote{Note that the synchrotron-self Compton
\citep{Waxman(1997),Ghisellini(1999)} have 
been ruled out by the high energy LAT observations \citep{Piran(2009), Zou(2009)}, while
external inverse Compton
\citep{Shemi(1994),Brainerd(1994),Shaviv(1995),Lazzati(2003)} require a strong enough external source of soft photons, which is not available in GRBs.} .

To examine the conditions for synchrotron to operate, we consider a general model that follows the spirit of \cite{Kumar(2008)}.
We don't make any specific assumptions concerning the details of the energy dissipation process or the particle acceleration mechanism.
Instead we assume that relativistic electrons and magnetic fields are at place within the emitting region,
that is moving relativistically towards the observer. We examine the parameter space in which these electrons emit the observed radiation
via synchrotron while satisfying all known limits on the observed prompt emission. 
We use a simple  ``single-zone'' calculations that do not take into account the
blending of radiation from different angles or from different emitting regions (reflecting matter moving at different Lorentz factors),
but for the purpose of obtaining the rough range of relevant parameters this is sufficient.

We characterize the conditions within the emitting region by six parameters:
the co-moving magnetic field strength, $B$, the number of relativistic emitters, $N_e$,
the ratio between the magnetic energy and the internal energy of the electrons, $\epsilon\equiv\epsilon_B/\epsilon_e$,
the bulk Lorentz factor of the emitting region, $\G$, the minimal electrons'
Lorentz factor in the source frame, $\gamma_m$ and the ratio between the shell crossing time
and the angular timescale, $k$.
We then characterize a single, ``typical", GRB pulse, which is
the building block of the GRB lightcurve, by three
basic quantities: the peak (sub-MeV) frequency, the peak flux and the duration. 
We compare the first two, the observed peak (sub-MeV) frequency and 
the peak flux with the predictions of the synchrotron model and
we use the observed duration of the pulse to limit the 
angular time scale \citep{Sari(1997)}. The synchrotron solution is incomplete without a determination of the accompanying synchrotron self-Compton (SSC) emission, which
influences the efficiency of the synchrotron emission. Furthermore,
the observations of the high energy SSC emission poses additional constraints on the emitting region. 
We describe, therefore, a self-consistent synchrotron self-Compton solution.

Three additional constraints should be taken into account \citep[see also][]{Daigne(2011)}.
First, energy considerations pose a strong lower limit on the efficiency. 
The energy of the observed sub-MeV flux, is huge and already highly constraining astrophysical models.
Thus, to avoid an ``energy crisis'', the emission process 
must be efficient and emit a significant fraction of the available energy in the sub-MeV band.
Second, the emission region cannot be optically thick.
Additionally, the LAT band (100 MeV- 300 GeV) GeV component is significantly weaker in most GRBs than 
the sub-MeV peak \citep{Granot(2009),Granot(2010),Beniamini(2011),Guetta(2011),Ackermann(2012)}.
We combine these three conditions and constrain the possible phase space in which synchrotron can produce the prompt emission.

The paper is organized as follows. We describe in \S \ref{basic} the basic concepts of the model, discussing the observations in \S \ref{obs} 
and the structure of the emitting region and the parameter phase space in \S \ref{emitreg}. In \S \ref{detmet} we describe the synchrotron equations
(\S \ref{synEq}), their solutions (\S \ref{synchsol}), the constraints from the accompanying SSC (\S \ref{sec:SSC}), the implications of
the results (\S \ref{sec:result}) and energy and efficiency constraints (\S \ref{energy_efficiency}). We continue with a discussion of specific issues
concerning the (i) expected spectral slopes and 
possibilities to alleviate the issue of the ``synchrotron line of death'' (\S \ref{specshape}) and (ii) the narrow $E_{peak}$ distribution (\S \ref{Epeak}).
In \S \ref{intshocks}, we deviate from the general philosophy of the paper and we discuss synchrotron emission within the context of the popular internal shocks framework.
We turn, in \S \ref{sec:paircreation}, to the implications of the pair creation limits from observed GeV emission for the synchrotron model.
In \S \ref{sec:CTA} we briefly discuss the possibilities of detecting GRBs with CTA, and their implications on the model at hand.

\section{The Global Picture}
\label{basic}

\subsection{The observations}
\label{obs}
Limits on the optical \citep{Roming(2006), Yost(2007), Klotz(2009)}, x-ray \citep{Obrien(2006)},
GeV \citep{Ando(2008),Guetta(2011),Beniamini(2011), Ackermann(2012)}
and TeV \citep{Atkins(2005), Albert(2006), Aharonian(2009)} emission during the prompt phase of GRBs,
demonstrate that the observed sub-MeV peak carries most of the GRB's energy.
The only (unlikely) possibilities that have not been ruled out are either an extremely strong and sharp peak between 10-100eV
or a peak at extremely high energies above the TeV range. 
Therefore, we associated the sub-MeV peak with the peak of the synchrotron emission.

We consider the typical observations after redshift corrections, i.e. in the host galaxy frame
which we denote as ``the source frame'' (as opposed to ``the observer frame'').
The basic observables are: the peak frequency: $h \nu_{peak}\equiv h \nu_{peak,obs}\times(1+Z)$ (300KeV),
the duration of a typical pulse, $t_p\equiv t_{p,obs}/(1+Z)$ (0.5 sec)
and the peak flux, $F_{\nu_{peak}}\equiv F_{\nu_{peak,obs}}/(1+Z)$ ($1.5\times10^{-26} \mbox{erg}/\mbox{cm}^2 \mbox{sec Hz}$).
With these definitions we remove the $(1+Z)$ dependence from the equations, and the only remaining
dependence on redshift is through the luminosity distance, relating the flux and luminosity.
All three quantities vary from pulse to pulse and the values in brackets denote the canonical values that we use here.
These three observations together with a typical luminosity distance, $d_L=2\times10^{28}$ cm (or $z=1$) correspond to an isotropic
equivalent luminosity of $L_{iso}\approx2\times10^{52}$erg/sec. It has been suggested that 
$E_{peak}$ and $L$ satisfy: $L_p=10^{52.43} ({E_{peak}(1+z)}/{355\mbox{KeV}})^{1.6}$erg/sec \citep{Yonetoku(2010)},
we therefore choose $\nu_{peak}$ and $L_p$ such that they are compatible with this relation.
In \S \ref{Epeak} we explore the dependence of the synchrotron model on this choice and look at 4 representative ``GRB types''.
An additional observation that we use is the limit on the total flux observed in the
LAT band (30 MeV-300 GeV), which is at most 0.13 of the GBM (8KeV-40MeV) flux
\citep{Beniamini(2011),Ando(2008),Guetta(2011),Ackermann(2012)}.

\subsection{The emitting region}
\label{emitreg}
The general model requires a relativistic jet, traveling with a Lorentz factor $\G$ with respect to
the host galaxy with a half opening angle of $\theta$, at a distance $R$ from the origin of the explosion
and with a kinetic energy $E_{tot}$ in the source frame. 
The source can be treated as spherically symmetric as long as $\G^{-1}<\theta$ which
is expected to hold during the prompt phase, in which $\G \geqslant 100$ \citep{Fenimore(1993),Woods(1995),Piran(1995),Sari(1999)}.
We therefore do not solve for $\theta$ and use isotropic equivalent quantities throughout the paper.
An upper limit on the emitting radius is given by the variability time scale,
$t_p\geqslant t_{ang}\equiv R/(2 c \G^2)$.
We define a dimensionless parameter $k$ such that $kR/G^2$ is the shell's width:
\begin{equation}
\label{t_cross}
t_{cross} \equiv \frac{kR}{2 c \G^2}=kt_{ang}. 
\end{equation}
With this definition, the pulse duration is:
\begin{equation}
\label{t_p}
t_{p} \equiv \frac{(k+1)R}{2 c \G^2}. 
\end{equation}
Clearly, for $k=1$ the pulse width is determined by angular spreading.

\begin{figure} [h]
\epsscale{0.5}
\plotone{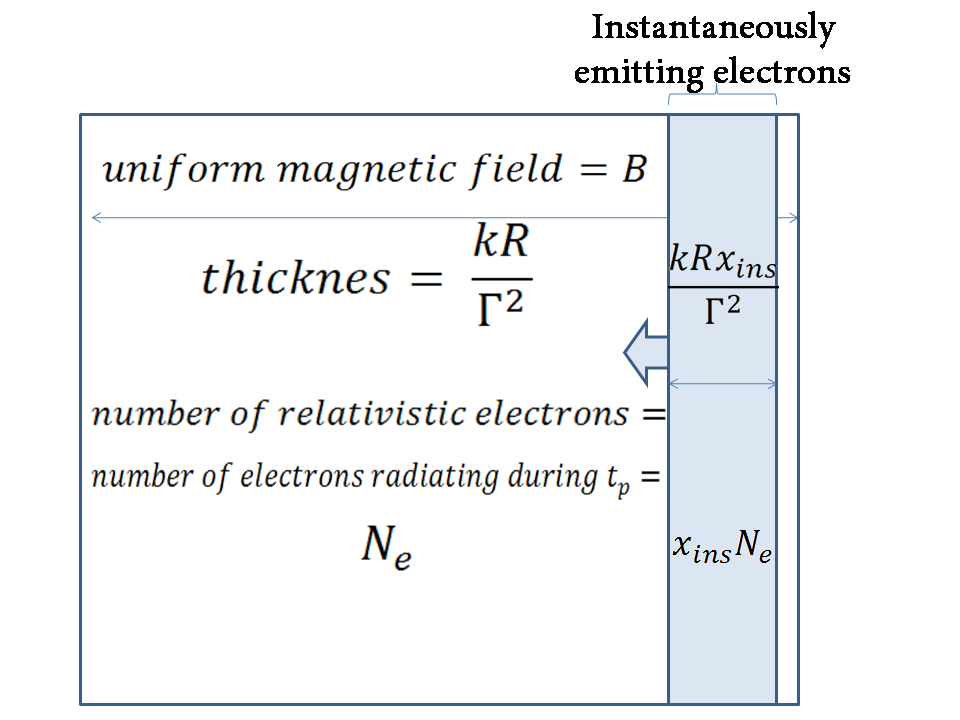}
\caption
{\small A schematic representation of the single zone model.
The model assumes that the physical conditions are uniform throughout the emitting region.
Due to the short cooling time by radiation (compared with the pulse duration) the instantaneously emitting particles
reside in a thinner slab than the overall population.} \label{fig:singlezone}
\end{figure}

We consider a single-zone model in which the different properties of the emitting region are constant
throughout the zone.
Notice that the single-zone approach cannot test spectral evolution, and becomes less accurate as $k$ increases.
Initially, the emitting region consists of either an electron-proton or an electron-positron plasma.
The numbers of protons and positrons are denoted by $N_p$ and $N_{e+}$ respectively.
Due to electrical neutrality, the number of electrons is $N_p+N_{e+}$.
After the gamma-rays are produced, annihilating high energy photons may result in extra electrons and positrons.
We denote
the number of pairs created in this way by $N_{pairs}$ (typically these particles are less energetic than the original
population of emitting particles).
Altogether, there are $N_{tot}=N_p+2N_{e+}+2N_{pairs}$ electrons and positrons in the flow.
the number of electrons is always larger than the number of positrons except for
the case of an initial pair dominated outflow, where they are equal.
Therefore, to simplify the presentation, we use hereafter "electrons" instead of
the more accurate but cumbersome "electrons and positrons".
In the fast cooling regime, the typical synchrotron cooling time of a relativistic electron
is shorter than the pulse duration and thus the instantaneously emitting particles are much less than the total number of
emitting particles during one pulse (see Fig.\ref{fig:singlezone}).
The emitting electrons are characterized by
$X_{ins}\leq1$ which is defined to be the ratio between the number of
instantaneously emitting electrons and the overall number of relativistic electrons emitting during the entire pulse
(A concise definition of all these particle numbers is given in table \ref{tbl:numbers}).

\begin{table}\small
\begin{tabular}{ l l}
\hline
Notation & Description\\
\hline
$N_{tot}$ & The total number of electrons (and positrons)\\
$N_{pairs}$ & The number of pairs created by annihilating high energy photons \footnotemark[1]\\
$N_p$ & The number of protons in the initial flow \\
$N_{e+}$ & The number of positrons in the initial flow \\
$N_e$ & The number of relativistic electrons (and positrons) radiating during one pulse\\
$x_{ins}$ & Ratio between instantaneously emitting electrons (and positrons) and $N_e$\\
\hline
\end{tabular}
 \caption{\small Different particle numbers used in the text.}
\footnotemark[1]{Excluding the original pairs that may reside in the initial flow and are accounted for by $N_{e+}$} 
\label{tbl:numbers}
\end{table}

The relativistic electrons have a power-law distribution of energies:
\begin{equation}
\frac{dN_e}{d\gamma}=C (\frac{\gamma}{\gamma_m})^{-p}, 
\end{equation}
(where C is a normalization constant) which holds for $\gamma_m<\gamma<\gamma_{Max}$. Based on theory, we expect $p$ to be of the order $p\sim 2.5$
\citep{Achterberg(2001), Bednarz(1998), Gallant(1999)}.
Indeed this result has been confirmed observationally both from the GRB prompt and afterglow phases \citep{Sari(1997b),Panaitescu(2001)}. 
$\gamma_{Max}$ arises, most likely, due to Synchrotron
losses at the energy where the acceleration time equals to the energy loss time \citep{de Jager(1996)}:
\begin{equation}
\label{synmax}
\gamma_{Max}=4 \times10^7 f B^{-1/2},
\end{equation}
where $f$ is a numerical constant of order unity which encompasses
the details of the acceleration process and $B$ is measured in Gauss.
In a shock-acceleration scenario this depends on the amount of time the particle spends in the
downstream and upstream regions \citep{Piran(2010), BDK(2011)}.
The total internal energy of the electrons is then (primes denote the co-moving frame):
\begin{equation}
\label{electronsE}
 E_e'=\frac{p-1}{p-2}\gamma_m N_em_e c^2\equiv\epsilon_e E_{int}',
\end{equation}
where $E_{int}'$ is the energy of the flow dissipated into internal energy
and the ratio of the total energy in these relativistic electrons to the total internal energy is denoted by $\epsilon_e$.
Notice that $\epsilon_e$ does not necessarily reflect the ratio of instantaneous energy of the relativistic electrons to the total energy, 
which could be much smaller.

The magnetic field, $B$, is assumed to be constant over the entire emitting region.
Its energy is:
\begin{equation}
\label{energymag}
E_B'=\frac{B^2}{8 \pi} 4\pi R^2 \frac{kR}{\G}\equiv \epsilon_B E_{int}'=\epsilon E_e',
\end{equation}
where $kR/\G$ is the thickness of the shell in the co-moving frame
and $\epsilon_B$ is the fraction of magnetic to internal energy.
In this case, for a Baryoninc dominated flow, $\epsilon_B+\epsilon_e\leqslant1$. Whereas,
in a Poyinting flux dominated jet, one can imagine a situation in which at first the energy is stored within the magnetic fields,
and only later it is dissipated and transferred to the electrons. Within the definitions we use here, this means that
$\epsilon_e+\epsilon_B \lesssim2$.

Overall, we find that six independent parameters define the synchrotron solution \footnote{\cite{Kumar(2008)}
set the pulse duration to be equal to
the angular time scale. We consider here a more general approach in which this is only a lower limit.} 
These are:
the co-moving magnetic field strength, $B$, the number of relativistic electrons, $N_e$,
the ratio between the magnetic energy and the kinetic energy of the electrons, $\epsilon\equiv\epsilon_B/\epsilon_e$,
the bulk Lorentz factor of the jet with respect to the GRB host galaxy, $\G$, the minimum electrons'
Lorentz factor in the source frame, $\gamma_m$, and the ratio between the shell crossing time
and the angular timescale, $k$.
The first three observables: the spectral peak, the peak luminosity and the pulse duration, described in \S \ref{obs},
reduce the number of free parameters from six to three, which we choose as: ($\epsilon_B/\epsilon_e,\G,k$).
The other constraints limit the allowed regions within
this parameter space.

\section{Detailed method and results}
\label{detmet}

\subsection{The synchrotron equations}
\label{synEq}
The typical synchrotron frequency and power produced by an electron with a Lorentz factor
$\gamma_e$ are: \citep{RybickiLightman(1979), Wijers_Galama(1999)}
\begin{equation}
\label{synch1}
\nu _{syn}=\G \gamma _e ^2 \frac{q_eB}{2 \pi m_e c },
\end{equation}
\begin{equation}
\label{sync2}
P (\gamma _e)=\frac{4}{3} \sigma _T c \G \gamma _e ^2 \frac{B^2}{8 \pi},
\end{equation}
where $\G$ is the Lorentz factor of the bulk and $\sigma _T$ is the Thomson cross-section.
When the electrons have a power-law distribution
of energies above some $\gamma_m$ there are two typical breaks in the spectrum.
The first is the synchrotron frequency of a typical electron, $\nu_m\equiv\nu _{syn}(\gamma_m)$.
The second is the cooling frequency, the frequency at which electrons cooling
on the pulse time-scale, radiate:
\begin{equation}
\label{coolfreq}
 \nu_c=\frac{18\pi q_e m_e c}{\sigma_T^2 B^3 \G t_p^2 {(1+Y_0)}^2},
\end{equation}
where $Y_0$ is the relative energy loss by IC as compared with synchrotron for $\gamma_e=\gamma_c$.

The peak of the synchrotron $\nu F_{\nu}$ is at $max[\nu_c,\nu_m]$ \citep{Sari(1998)}.
We identify this peak with the observed sub-MeV peak.
For $\nu _c<\nu _m$, the "typical" electrons are fast cooling and rapidly dispose of
all their kinetic energy, while for $\nu _c>\nu _m$ the electrons are slow cooling and the typical electrons do not emit
all their energy which is eventually lost by adiabatic cooling.
We consider fast cooling solutions (or marginally fast) for the prompt emission.

The maximal spectral flux, at $min[\nu_c,\nu_m]$ is given by combining Eqns. \ref{synch1}, \ref{sync2}:
\begin{equation}
\label{synch2}
F _{\nu ,max} =\frac{m_e c^2 \sigma _T \G B N_e}{3 q_e 4 \pi d_L^2}.
\end{equation}
For $\nu _c<\nu _m$, this is the flux at $\nu_c$ and the peak of $F_\nu$.
However, in the fast cooling regime the flux between $\nu_c$ and $\nu_m$ varies as $F_\nu \propto \nu^{-1/2}$
and the peak of $\nu F_\nu$ is at $\nu_m$:
\begin{equation}
\label{synch3}
F _{\nu ,m} = F _{\nu ,max} (\frac{\nu_m}{\nu_c})^{-1/2}=\frac{m_e c^2 \sigma _T \G B x_{ins}N_e}{3 q_e 4 \pi d_L^2(1+Y_0)} ,
\end{equation}
where the factor $x_{ins}/(1+Y_0)\equiv(\nu_c/\nu_m)^{1/2}$ accounts for the fact that only $x_{ins}$ of the relativistic electrons are instantaneously
emitting and only $1/(1+Y_0)$ of their energy loss is by synchrotron (the rest is by the accompanying SSC).
In the fast cooling regime, the electrons cool on a shorter timescale than the pulse duration
and several consecutive cooling processes may be needed 
to account for the comparatively long pulse duration.
Eq. \ref{synch3} can be understood in terms of the cooling time of the electrons, $t_c$,
the time it takes the electrons to radiate (by synchrotron and IC) their kinetic energy:
\begin{equation}
\label{cooltime}
t_c(\gamma_e)=\frac{\G \gamma_e m_e c^2}{\frac{4}{3} c \sigma_T \gamma_e^2 \G^2 \frac{B^2}{8\pi}(1+Y_0)}.
\end{equation}
$t_c(\gamma_e)$ is declining as a function of $\gamma_e$. Therefore it is maximal for $\gamma_m$, and after a time
$t_c(\gamma_m)$ all of the electrons have cooled significantly.
Combining Eqs. \ref{synch1}, \ref{coolfreq} and \ref{cooltime} gives: $t_c(\gamma_m)/t_p=(\nu_c/\nu_m)^{1/2}$.
It follows that: $x_{ins}/(1+Y_0)=t_c(\gamma_m)/t_p$, which tells us that indeed only the instantaneously synchrotron emitting
electrons contribute to the peak of $\nu F_{\nu}$.

Given the spectrum, the source must be optically thin for scatterings:
\begin{equation}
\label{tau1}
\tau=\frac{N_{tot} \sigma_T}{4 \pi R^2}\lesssim1,
\end{equation}
where $N_{tot}=N_e+2N_{pairs}$ is the total number of electrons and positrons in the flow. $N_{tot}$ is composed of two components.
The first, $N_e$, is calculated from the peak flux.
Notice that in principle there could be additional non-relativistic electrons in the initial flow that although they don't contribute
to the radiation, they can be important due to their effect on the optical depth.
We return to this point in \S \ref{synchsol}.
The second, $2N_{pairs}$, is the electron-positron pairs created by annihilation of the high energy photons
\footnote{The pairs are created with energies of order half that of the high energy photon. This causes the pairs to have a steep power law of random Lorentz factors
extending up to very high energies. These pairs radiate by synchrotron and contribute mainly at the optical band. Their radiation in the sub-MeV band is
sub-dominant.}.
$N_{pairs}$ is derived by considering the optical depth for pair creation (\cite{Lithwick(2001)}):
\begin{equation}
\label{taugg}
\tau_{\gamma,\gamma}(\nu)=\frac{\frac{11}{180}\sigma_T n_{>\nu,an}}{4 \pi R^2}\lesssim1 
\end{equation}
where $n_{>\nu,an}$ is the number of photons in the source with enough energy to annihilate
the photon with frequency, $\nu$.
The frequency $\nu_{thick}(\G)$, is defined as the (source frame) photon frequency at which $\tau_{\gamma,\gamma}(\nu_{thick})=1$.
Therefore, $N_{pairs}$ equals the number of photons above $\nu_{thick}(\G)$.

A more detailed calculation of the pair creation opacity has to take care of the details of the radiation field in the source \citep{Granot(2008),Hascoet(2012)}.
There could in principle be a numerical factor in Eq. \ref{taugg} of order $\sim 0.01-0.1$ \citep{Granot(2008),Hascoet(2012),Vurm(2012)}
This in turn, would lower the minimum requirement on $\G$. However: $\G \propto \tau^{\frac{1}{6}}$,
which means that the change in the lower limit of $\G$ in the exact calculation, will only be of order 0.5-0.7.

\subsection{A synchrotron solution for the MeV peak}
\label{synchsol}
We solve the synchrotron Eqns. described in \S \ref{synEq} and reduce the 6D parameter space described in \S \ref{emitreg} to a 3D
parameter space defined by ($\epsilon,\G,k$). We then consider three constraints that limit the allowed regions within this parameter space.
In \S \ref{sec:paircreation} we discus further constraints on this parameter space that arise from pair creation opacity limits.

Solving Eqns. \ref{t_p}, \ref{energymag}, \ref{synch1}, \ref{coolfreq} and \ref{synch3} we find:
\begin{equation}
\label{mag1}
B=2\times10^4 (1+Y_0)^{1/2}\G_2^{-3} (\epsilon)^{1/2} k^{-1/2} (k+1)^{3/2} \nu_{p,300}^{1/2}t_{p,.3}^{-1} F_{-26.2}^{1/2} d_{28} \mbox{ Gauss},
\end{equation}
\begin{equation}
\label{xins}
x_{ins}=1.1 \times 10^{-5} (1+Y_0)^{-7/4}\G_2^{4} (\epsilon)^{-3/4} k^{3/4} (k+1)^{-9/4} \nu_{p,300}^{-5/4} t_{p,.3}^{1/2} F_{-26.2}^{-3/4} d_{28}^{3/2} ,
\end{equation}
\begin{equation}
 \label{N_{rel}}
N_e=1.8\times10^{52} (1+Y_0)^{5/4}\G_2^{-2} (\epsilon)^{1/4} k^{-1/4} (k+1)^{3/4} \nu_{p,300}^{3/4} t_{p,.3}^{1/2} F_{-26.2}^{5/4} d_{28}^{5/2} ,
\end{equation}
\begin{equation}
\label{R}
R=3\times10^{14} \G_2^2 (k+1)^{-1}t_{p,.3} \mbox{ cm},
\end{equation}
\begin{equation}
\label{gamma}
\gamma_m=3600 (1+Y_0)^{-1/4}\G_2 (\epsilon)^{-1/4} k^{1/4} (k+1)^{-3/4} \nu_{p,300}^{1/4} t_{p,.3}^{1/2} F_{-26.2}^{-1/4} d_{28}^{-1/2},
\end{equation}
\begin{equation}
\label{gammaMax}
\gamma_{Max}=2.8\times10^5 (1+Y_0)^{-1/4}\G_2^{3/2} (\epsilon)^{-1/4} f k^{1/4} (k+1)^{-3/4} \nu_{p,300}^{-1/4} t_{p,.3}^{1/2} F_{-26.2}^{-1/4} d_{28}^{-1/2},
\end{equation}
where $\G_2\equiv \G/100$, $\nu_{p,300}\equiv h \nu _{peak}/300$KeV,
$F_{-26.2} \equiv F _{\nu ,m} /(1.5\times10^{-26}\mbox{erg sec}^{-1} \mbox{Hz} ^{-1} \mbox{cm}^{-2})$,
$t_{p,.3}\equiv t_p/0.5$sec and $d_{28}\equiv d_L/10^{28}$cm.

We proceed to consider the limits on the 
parameter space ($\epsilon,\G,k$) that arise from (i) efficiency (fast cooling), (ii) the size of the emitting radius and (iii) the opacity. 
 
Efficiency and fast cooling require $\nu_c<\nu_m$ (\S \ref{synEq}), which yields: 
\begin{equation}
\label{coolm}
 \G_2<17 (1+Y_0)^{7/16}(\epsilon)^{3/16} k^{-3/16} (k+1)^{9/16} \nu_{p,300}^{5/16} t_{p,.3}^{-1/8}F_{-26.2}^{3/16} d_{28}^{3/8} .
\end{equation}

The emitting radius must be smaller than $R_{dec}$, the deceleration radius \citep{Panaitescu(2004),Kumar(2008), Zou(2009)}.
The latter depends on the circum-burst density profile: a wind with $n=(A/m_p)r^{-2}$ or a constant ISM with $n\sim n_0$.
The limit on the emitting region can be written, in these two cases, as:
\begin{equation}
\label{decRad}
R< R_{dec} = 
\left\{
  \begin{array}{l l}
    2.6\times10^{17}E_{54}^{1/3}n_0^{-1/3}\G_2^{-2/3}\mbox{max}(1,0.01n_0^{1/7}T_2^{3/7}E_{54}^{-1/7}\G_2^{-2/3}) \mbox{cm}, & \quad \text{constant ISM}\\
    1.8\times10^{16}E_{54}A_*^{-1}\G_2^{-2}\mbox{max}(1,0.36A_*^{1/2}T_2^{1/2}E_{54}^{-1/2}\G_2^2) \mbox{cm}, & \quad 
    \text{wind environment}\\
  \end{array} \right.
\end{equation}
where $E_{54} \equiv E_{iso}/10^{54}$erg is the total (isotropic) energy of the burst (and not just the energy emitted in one pulse),
$T_2\equiv T/100$sec is the total duration of the burst, $n_0$ is the particle density in $\mbox{cm}^{-3}$ and $A_*\equiv A/(5\times10^{11}\mbox{g/cm})$.

The flow is optically thin for scattering.
Typically, we find that $N_{tot}$ is dominated by the pairs created within the flow by the annihilation of the high energy part of the spectrum.
The main contribution to the number of photons available for pair creation is from IC scatterings of synchrotron
photons to energies above the pair creation threshold.
Solving Eqns. \ref{tau1}, \ref{taugg} and \ref{gamma} we obtain:
\begin{equation}
\label{tau2}
\tau=8.7  (1+Y_0)^{3/8}\G_2^{-6}(\epsilon)^{-1/8} k^{-3/8} (k+1)^{25/8}\nu_{p,300}^{1/8} t_{p,.3}^{-7/4} F_{-26.2}^{11/8} d_{28}^{11/4}.
\end{equation}
The criterion $\tau<1$ leads to a lower limit on $\G$.
It should be noted that in many models one may expect a large amount of passive (i.e. non relativistic) electrons in the flow \citep{Bosnjak(2009)}.
In this case the optical depth may become dominated by these electrons. We discuss this option in \S \ref{intshocks}.

So far we discussed only the synchrotron signal itself, and we haven’t yet addressed the SSC contribution.
We now turn to describe the SSC emission, and explore its effect on the solutions.
\subsection{The SSC component}
\label{sec:SSC}

Synchrotron is essentially accompanied by SSC.
We describe here a single-zone analytic solution of this component which is expected to reproduce the general
characteristics of the true solution. However, we note that a full description of the up-scattered flux can only be done with
more detailed calculations \citep{Nakar(2009),Bosnjak(2009)} and those might differ somewhat from the simplified analysis.
If a SSC signal is detected, we can measure both the peak frequency and the flux of the SSC component.
In this case, we obtain two additional equations that allow us to determine the parameters up to a single free parameter
which we choose as $k$.
This may be the situation in the future, when CTA begins observing in the very high energy range. We return to this intriguing possibility in \S \ref{sec:CTA}. But even if the SSC signal 
is not detected (as is the current observational situation) we can use the non-detection as a limit on the parameter space. Note that the signal could be undetected 
 either because it  is very weak or because it peaks at sufficiently high frequencies and cannot be observed by current detectors.

The typical SSC photon frequency is given by:
\begin{equation}
 \nu_{SSC}=\gamma_m^2 \nu_m .
\end{equation}
The IC process is suppressed by the Klein-Nishina effect and the photons are up-scattered only to $\nu_{KN}$, if the up-scattered photon's energy in the electron's rest frame exceeds the electron rest mass energy:
\begin{equation} 
\label{KNnu}
h \nu_{KN}\equiv \G \gamma_m m_e c^2<\gamma_m ^2 h\nu_m .
\end{equation}
This condition is satisfied if:
\begin{equation}
\epsilon<1.8\times10^5 k^{-1}(1\!+\!k)^3\nu_{p,300}^{5}t_{p,.3}^2 F_{-26.2}^{-1}d_{28}^{-2}.
\end{equation}
Thus, for typical parameters the up-scattered photons are most likely in the KN regime.
We introduce the parameter $\zeta_{KN}$ \citep{Ando(2008)} to take the KN effect into account: 
\begin{equation}
\label{zetaKN}
\zeta_{KN} \sim \begin{cases} (\frac{\nu_{SSC}}{\nu_{KN}})^{-\frac{1}{2}}=(\frac{\gamma_m h \nu_m}{\G m_e c^2})^{-\frac{1}{2}} = 0.22 (1+Y_0)^{\frac{1}{8}}(\epsilon)^{\frac{1}{8}} k^{-\frac{1}{8}}(1\!+\!k)^{\frac{3}{8}}
\nu_{p,300}^{-\frac{5}{8}}t_{p,.3}^{-\frac{1}{4}}F_{-26.2}^{\frac{1}{8}}d_{28}^{\frac{1}{4}} , & \mbox{if } \frac{\nu_{SSC}}{\nu_{KN}}>1 \\ 1, & \mbox{else}  .\end{cases}
\end{equation}
 
The total SSC flux (integrated over frequencies), $F_{SSC}$, is related to the total flux ascosiated with the synchrotron peak, $\sim\nu_m F_{\nu,m}$,
by:
\begin{equation}
\label{KNratio}
\frac{F_{SSC}}{\nu_m F_{\nu,m}}\equiv Y_0=Y \zeta _{KN}, 
\end{equation}
where:
\begin{equation}
\label{Ycases}
Y = \begin{cases} (\frac{\epsilon_e}{\epsilon_B})^{1/2}, & \mbox{if } \epsilon_e >\epsilon_B \\ \frac{\epsilon_e}{\epsilon_B}, & \mbox{if } \epsilon_e <\epsilon_B \end{cases},
\end{equation}
is the Compton parameter \citep{Sari(1996)} .

This yields:
\begin{equation}
\label{Y0}
Y_0\!=\!\begin{cases}
0.22\epsilon^{-\frac{7}{8}} k^{-\frac{1}{8}}(1\!+\!k)^{\frac{3}{8}}\nu_{p,300}^{-\frac{5}{8}}t_{p,.3}^{-\frac{1}{4}}F_{-26.2}^{\frac{1}{8}}d_{28}^{\frac{1}{4}}, &\! \mbox{if }  \epsilon\!>\!1 (Y_0\!<\!Y\!<\!1)\\
0.22\epsilon^{-\frac{3}{8}} k^{-\frac{1}{8}}(1\!+\!k)^{\frac{3}{8}}\nu_{p,300}^{-\frac{5}{8}}t_{p,.3}^{-\frac{1}{4}}F_{-26.2}^{\frac{1}{8}}d_{28}^{\frac{1}{4}}, &\! \mbox{if }
1\!>\!\epsilon>\!1.7\!\times\!10^{-2} k^{-\frac{1}{3}}(1\!+\!k)\nu_{p,300}^{-\frac{5}{3}}t_{p,.3}^{-\frac{2}{3}} F_{-26.2}^{\frac{1}{3}}d_{28}^{\frac{2}{3}} (Y_0\!<\!1\!<\!Y)\\
0.17\epsilon^{-\frac{3}{7}} k^{-\frac{1}{7}}(1\!+\!k)^{\frac{3}{7}}\nu_{p,300}^{-\frac{5}{7}}t_{p,.3}^{-\frac{2}{7}}F_{-26.2}^{\frac{1}{7}}d_{28}^{\frac{2}{7}}, &\! \mbox{else }(1\!<\!Y_0\!<\!Y)  \end{cases}
\end{equation}

As mentioned above, a possible future detection of an SSC component above the LAT range can, in principle, measure: $\nu_{KN}$ and  $F_{SSC}$. This provides two extra equations, allowing us to
solve for $\G$ and $\epsilon$. Using Eqns. \ref{KNratio} \ref{Ycases} we can write:
\begin{equation}
\epsilon\!=\!\begin{cases}
0.17 Y_{0,-1}^{-\frac{8}{7}} k^{-\frac{1}{7}}(1\!+\!k)^{\frac{3}{7}} \nu_{p,300}^{-\frac{5}{7}}t_{p,.3}^{-\frac{2}{7}}F_{-26.2}^{\frac{1}{7}}d_{28}^{\frac{2}{7}}, & \mbox{if }  Y_0\!<\!Y\!<\!1,\\
1.7 \times 10^{-2} Y_{0,-1}^{-\frac{8}{3}} k^{-\frac{1}{3}}(1\!+\!k) \nu_{p,300}^{-\frac{5}{3}}t_{p,.3}^{-\frac{2}{3}}F_{-26.2}^{\frac{1}{3}}d_{28}^{\frac{2}{3}}, & \mbox{if }
Y_0\!<\!1\!<\!Y,\\
1.7 \times 10^{-2} Y_{0,-1}^{-\frac{7}{3}} k^{-\frac{1}{3}}(1\!+\!k) \nu_{p,300}^{-\frac{5}{3}}t_{p,.3}^{-\frac{2}{3}}F_{-26.2}^{\frac{1}{3}}d_{28}^{\frac{2}{3}}, & \mbox{else }(1\!<\!Y_0\!<\!Y);  \end{cases}
\end{equation}
and
\begin{equation}
\G\!=\!\begin{cases}
80 Y_{0,-1}^{-\frac{1}{7}} k^{-\frac{1}{7}}(1\!+\!k)^{\frac{3}{7}} \nu_{p,300}^{-\frac{3}{14}}t_{p,.3}^{-\frac{2}{7}}F_{-26.2}^{\frac{1}{7}}d_{28}^{\frac{2}{7}}(\frac{h\nu_{KN}}{100GeV})^{\frac{1}{2}}, & \mbox{if }  Y_0\!<\!Y\!<\!1,\\
60 Y_{0,-1}^{-\frac{1}{3}} k^{-\frac{1}{6}}(1\!+\!k)^{\frac{1}{2}} \nu_{p,300}^{-\frac{1}{3}}t_{p,.3}^{-\frac{1}{3}}F_{-26.2}^{\frac{1}{6}}d_{28}^{\frac{1}{3}}(\frac{h\nu_{KN}}{100GeV})^{\frac{1}{2}}, & \mbox{if }
Y_0\!<\!1\!<\!Y,\\
60 Y_{0,-1}^{-\frac{1}{6}}k^{-\frac{1}{6}}(1\!+\!k)^{\frac{1}{2}} \nu_{p,300}^{-\frac{1}{3}}t_{p,.3}^{-\frac{1}{3}}F_{-26.2}^{\frac{1}{6}}d_{28}^{\frac{1}{3}}(\frac{h\nu_{KN}}{100GeV})^{\frac{1}{2}}, & \mbox{else }(1\!<\!Y_0\!<\!Y). \end{cases}
\end{equation}

In addition, detection of a SSC signal above tens of GeV, implies that the spectral breaks observed by the LAT
are not due to an optical depth.
If the particles are shock accelerated then
these breaks would be associated with the maximal synchrotron frequency: $\nu_{syn,max}=4.5\times 10^{23}f^2\G_2$Hz\citep{de Jager(1996)}
where $f$ depends on the acceleration mechanism, and is expected to be of order unity \citep{Piran(2010), BDK(2011)}.
If we  can also observe $\nu_{syn,max}$ we obtain another equation that would allow us to solve for $k$
and either obtain the full solution or rule it out if no solution is found:
\begin{equation}
\label{nusynmax}
\G=100 f^{-2} (h\nu_{syn,max}/1GeV)
\end{equation}
\begin{equation}
\epsilon\!=\!\begin{cases}
0.2 f^{-2} Y_{0,-1}^{-1} (\frac{h\nu_{syn,max}}{1GeV})(\frac{h\nu_{KN}}{100GeV})^{-\frac{1}{2}} \nu_{p,300}^{-\frac{1}{2}}, & \mbox{if }   Y_0\!<\!Y\!<\!1\\
4.5 \times 10^{-2} f^{-4} Y_{0,-1}^{-2} (\frac{h\nu_{syn,max}}{1GeV})^{2}(\frac{h\nu_{KN}}{100GeV})^{-1} \nu_{p,300}^{-1}, & \mbox{if }
Y_0\!<\!1\!<\!Y\\
4 \times 10^{-2} f^{-4} Y_{0,-1}^{-2} (\frac{h\nu_{syn,max}}{1GeV})^{2}(\frac{h\nu_{KN}}{100GeV})^{-1} \nu_{p,300}^{-1}, & \mbox{else }(1\!<\!Y_0\!<\!Y), \end{cases}
\end{equation}

\begin{equation}
k\!=\!\begin{cases}
2.2 f^{-7} Y_{0,-1}^{\frac{1}{2}} (\frac{h\nu_{syn,max}}{1GeV})^{\frac{7}{2}}(\frac{h\nu_{KN}}{100GeV})^{-\frac{7}{4}} \nu_{p,300}^{\frac{3}{4}}t_{p,.3} F_{-26.2}^{-\frac{1}{2}} d_{28}^{-1}, & \mbox{if }  Y_0\!<\!Y\!<\!1\\
4  f^{-6} Y_{0,-1} (\frac{h\nu_{syn,max}}{1GeV})^{3}(\frac{h\nu_{KN}}{100GeV})^{-\frac{3}{2}} \nu_{p,300}t_{p,.3} F_{-26.2}^{-\frac{1}{2}} d_{28}^{-1}, & \mbox{if } 
Y_0\!<\!1\!<\!Y\\
4.7  f^{-6} Y_{0,-1} (\frac{h\nu_{syn,max}}{1GeV})^{3}(\frac{h\nu_{KN}}{100GeV})^{-\frac{3}{2}} \nu_{p,300}t_{p,.3} F_{-26.2}^{-\frac{1}{2}} d_{28}^{-1}, & \mbox{else }(1\!<\!Y_0\!<\!Y).  \end{cases}
\end{equation}

\begin{figure}
\plotone{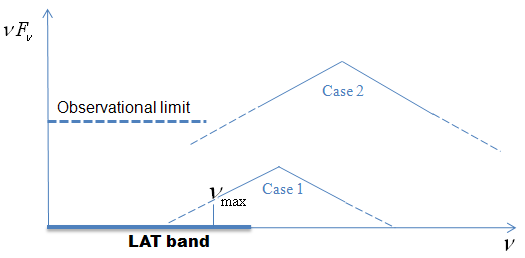}
\caption
{\small Two schematic possibilities allowing the SSC component to be compatible with LAT observations.} \label{fig:GeVopt}
\end{figure}

As opposed to similar observations in Blazars, current observations in the GeV range do not detect a
significant high energy component in typical GRBs.
Current detections of a GeV component are at the level of 0.03 of the MeV emission and upper limits for most other bursts are at the level of 0.1. In  a few cases (e.g.  GRB 090926A \citep{Ackermann(2011)}) an additional high energy component has
been detected, and may be associated with SSC,
however, even in this case one does not see a clear GeV peak, like the one observed in Blazars. 
In this case, the upper limits on the SSC signal further constrain the possible parameter space.
First, the photon with largest observed frequency, $\nu_{max}$ must satisfy: $\nu_{max}\leqslant\nu_{thick}$,
where the latter is the frequency at which the flow becomes optically thick to pair creation.
Above $\nu_{max}$, the photons could (but do not have to) be optically thick for pair creation.
The condition $\tau_{\gamma,\gamma}(\nu_{max})\leqslant1$ leads to lower limits on $\G$.
However, this condition is somewhat model dependent \citep{Granot(2008), Zou(2009), Zou(2011), Hascoet(2012)}, and the limits on $\G$ from this consideration may be alleviated in
case of a two zone model. We therefore separate the discussion of this limit from the main analysis, and return to it later on in \S \ref{sec:paircreation}.
Second, \cite{Beniamini(2011)} have shown that for a typical GBM burst the total flux which is observed in the
LAT band (30 MeV-300 GeV) is at most 0.13 of the GBM (8KeV-40MeV) flux of the same burst (see also \citealt{Ando(2008)}, \citealt{Guetta(2011)}).
More recent studies, using extra noise cuts applied by the LAT team \citep{Ackermann(2012)} give more constraining upper limits which are lower
by a factor $\sim 3$.
We define $\eta_{LAT}$ as the upper limit on the fractional LAT flux: ${F_{LAT}}/{F_{\nu,m}}\leqslant0.1 \eta_{LAT-1}$ (where $\eta_{LAT-1}\equiv\eta_{LAT}/0.1$).
Below $\nu_{max}$, the photons are necessarily optically thin for pair creation and therefore in order for
the LAT signal to be sufficiently low, either the total up-scattered flux is low
(less than $\eta_{LAT}$ of the GBM flux), or most of the up-scattered flux is at $\nu>\nu_{max}$ 
and only a small fraction can be observed (for an illustration of these possibilities see Fig. \ref{fig:GeVopt}).
This leads to lower limits on $\G$ and on $\epsilon$.

Let $\alpha$ denote the photon spectral index below $\nu_{peak}$, i.e. $N_{\nu} \propto \nu^{\alpha}$.
Most GRBs are phenomenologically well fit by a Band function, and for these $\alpha \approx -1$.
Defining $\zeta_W$ as the fraction of up-scattered flux observed in the LAT window, relative to the total up-scattered flux,
the ratio between the expected flux up to $\nu_{max}$ and the sub-MeV flux, is:
 \begin{equation}
\label{etaIC}
\frac{F_{LAT}}{\nu_m F_{\nu,m}} =Y_0 \zeta _{W}.
\end{equation}
If $\nu_{KN}>\nu_{max}$ and if $\nu F_{\nu}$ is rising below $\nu_{KN}$
(as happens for $\alpha \approx -1$) then the total flux observed in the LAT window is dominated by $\nu_{max}$.
$\zeta_W$ can be written as \citep{Ando(2008)}:
\begin{equation}
\label{xiW}
\zeta _{W} \approx \begin{cases} 1 , & \mbox{if }  min[\nu _{SSC},\nu_{KN}]<\nu_{max}\\ (\frac{ min[\nu _{SSC},\nu_{KN}]}{\nu_{max}})^{-\alpha-2},
& \mbox{else.} \end{cases}
\end{equation}
Plugging in $\nu_{KN}$ (i.e. assuming $\nu_{KN}<\nu_{SSC}$) yields:
\begin{equation}
\label{xiW1}
\zeta _{W} = \begin{cases} 1,\;\;\;\;\;\;\;\; \;\;\;\;
\text{ if }180\nu_{max,1}^{-1}\G_2^2 (\epsilon)^{\frac{-1}{4}}k^{1/4}(1\!+\!k)^{-3/4} \nu_{p,300}^{1/4}t_{p,.3}^{1/2}F_{-26.2}^{-1/4}d_{28}^{-1/2}<1
\\ (180\nu_{max,1}^{-1}\G_2^2 (\epsilon)^{\frac{-1}{4}}k^{1/4}(1\!+\!k)^{-3/4} \nu_{p,300}^{1/4}t_{p,.3}^{1/2}F_{-26.2}^{-1/4}d_{28}^{-1/2})^{-\alpha-2}
\text{ else}
 \end{cases}
\end{equation}
where $h\nu_{max,1}\equiv \frac{\nu_{max}}{1 GeV}$.

We require that the SSC contribution will be sufficiently low to agree with the LAT observations.
We separate the possible solutions to two cases.

If $\epsilon>2.4 k^{-1/7}(1+k)^{3/7} \nu_{p,300}^{-5/7}t_{p,.3}^{-2/7}F_{-26.2}^{1/7}d_{28}^{2/7}\eta_{LAT-1}^{-8/7}$,
then, $\frac{F_{SSC}}{\nu_m F_{\nu,m}}=Y_0<0.1\eta_{LAT-1}$
and the total up-scattered flux (and not just the fraction within the considered band) is below the observed limit. 
Otherwise, the total ratio of up-scattered to synchrotron flux
is larger than $0.1\eta_{LAT-1}$, but it peaks at $\nu_{KN}\gg\nu_{max}$ and thus may be partially absorbed.
As $\G$ increases, the up-scattered flux peaks at higher frequencies. This means, that for large enough values of $\G$,
the up-scattered tail below $\nu_{max}$ is small enough that it is compatible with the observations.
Therefore, in this regime of $\epsilon$, one can obtain a lower limit on $\G$ in terms of $\eta_{LAT}$
arising from the requirement that the total up-scattered flux up to $\nu{max}$ be less than the flux observed by LAT:
\begin{equation}
\label{Glimit}
 \G_2>10^{-\frac{2.25\alpha-4.27}{2\alpha+4}} \eta_{LAT-1}^{\frac{-1}{2\alpha+4}} \nu_{max,1}^{1/2}\epsilon^{\frac{2\alpha+1}{8(2\alpha+4)}}
k^{-\frac{2\alpha+5}{8(2\alpha+4)}} (1\!+\!k)^{\frac{6\alpha+15}{8(2\alpha+4)}} \nu_{p,300}^{-\frac{2\alpha+9}{8(2\alpha+4)}}t_{p,.3}^{-\frac{2\alpha+5}{4(2\alpha+4)}}F_{-26.2}^{\frac{2\alpha+5}{8(2\alpha+4)}}.
\end{equation}
Notice that these are conservative limits as we only use the spectral range below $\nu_{max}$ where we know the conditions are optically thin.
One must bear in mind, that efficiency considerations limit the amount of energy that may be carried by
the up-scattered flux by virtue of Eq. \ref{etatot}. This limits $\epsilon$ but not $\G$.

\subsection{Results}
\label{sec:result}

\begin{figure}
\centering
\epsscale{1.1}
\plottwo{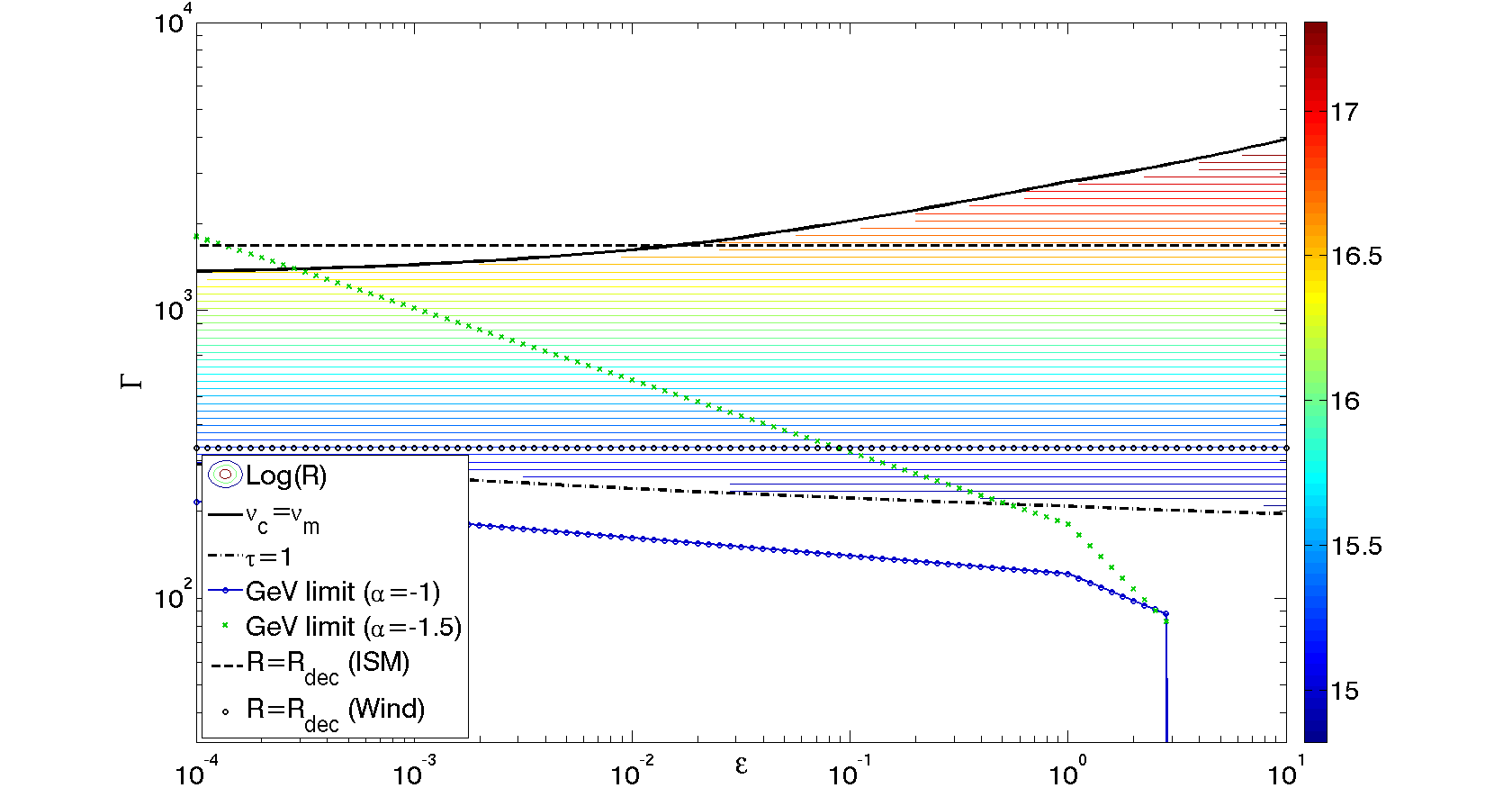}{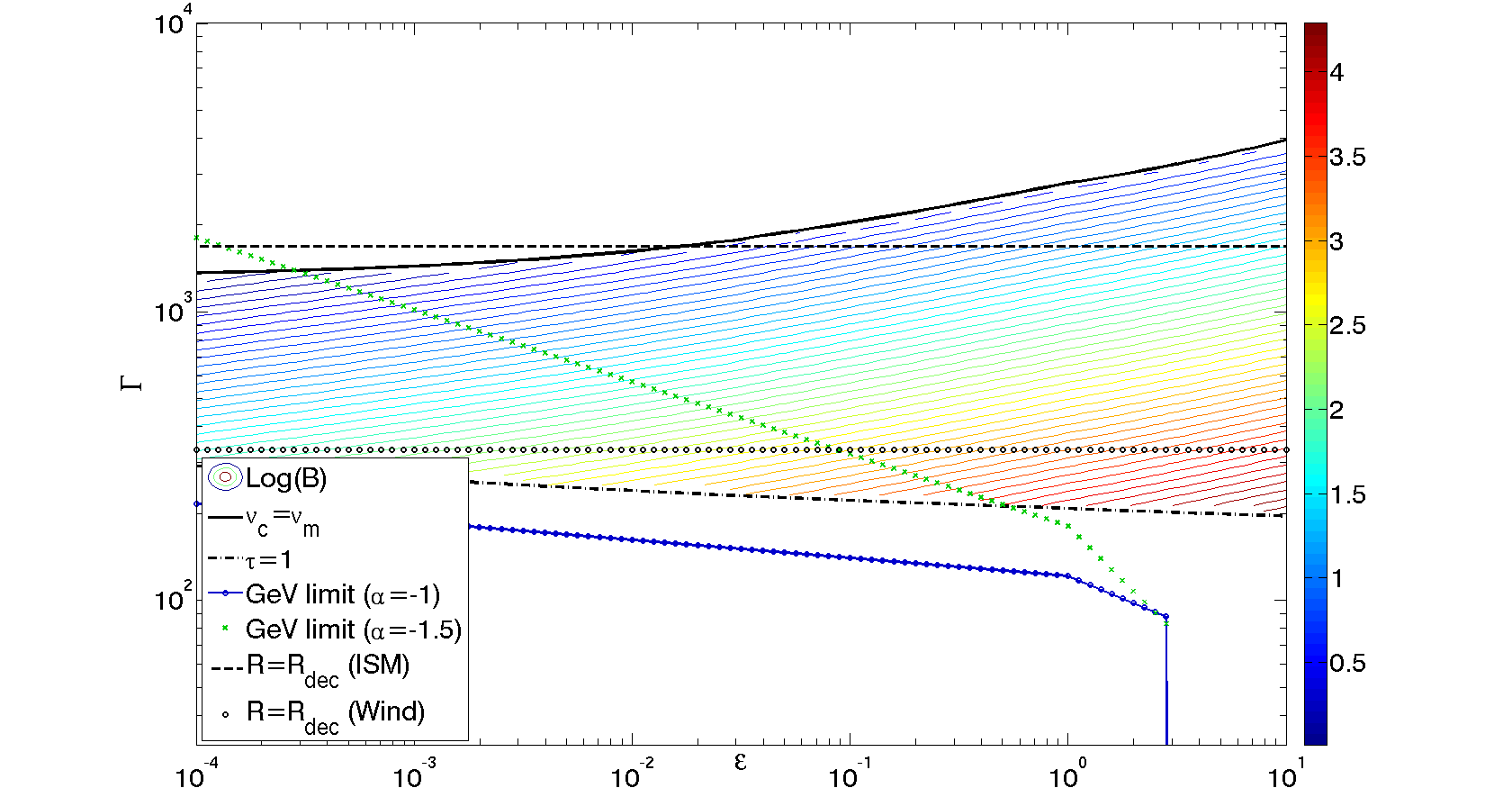}\\
\plottwo{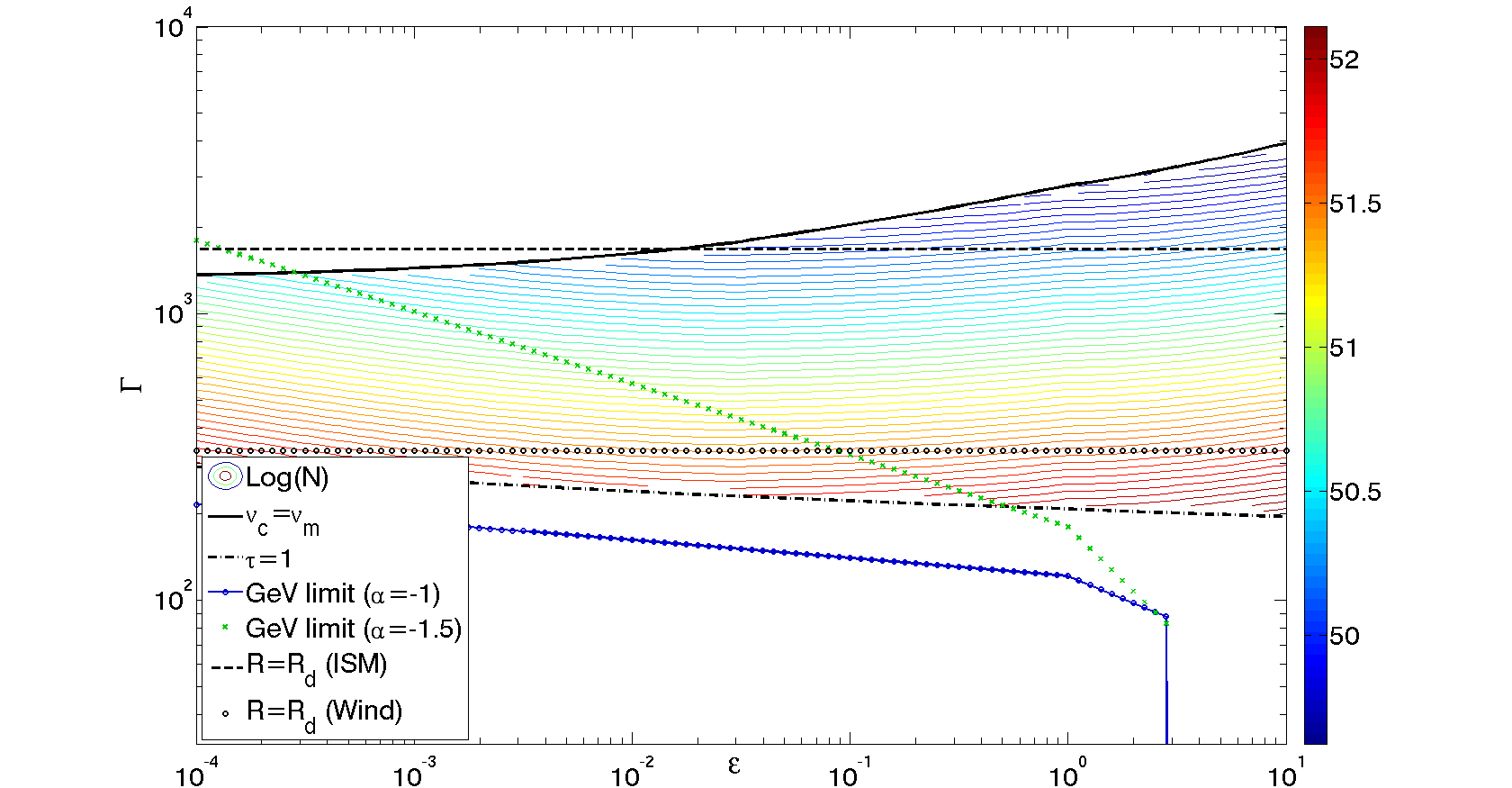}{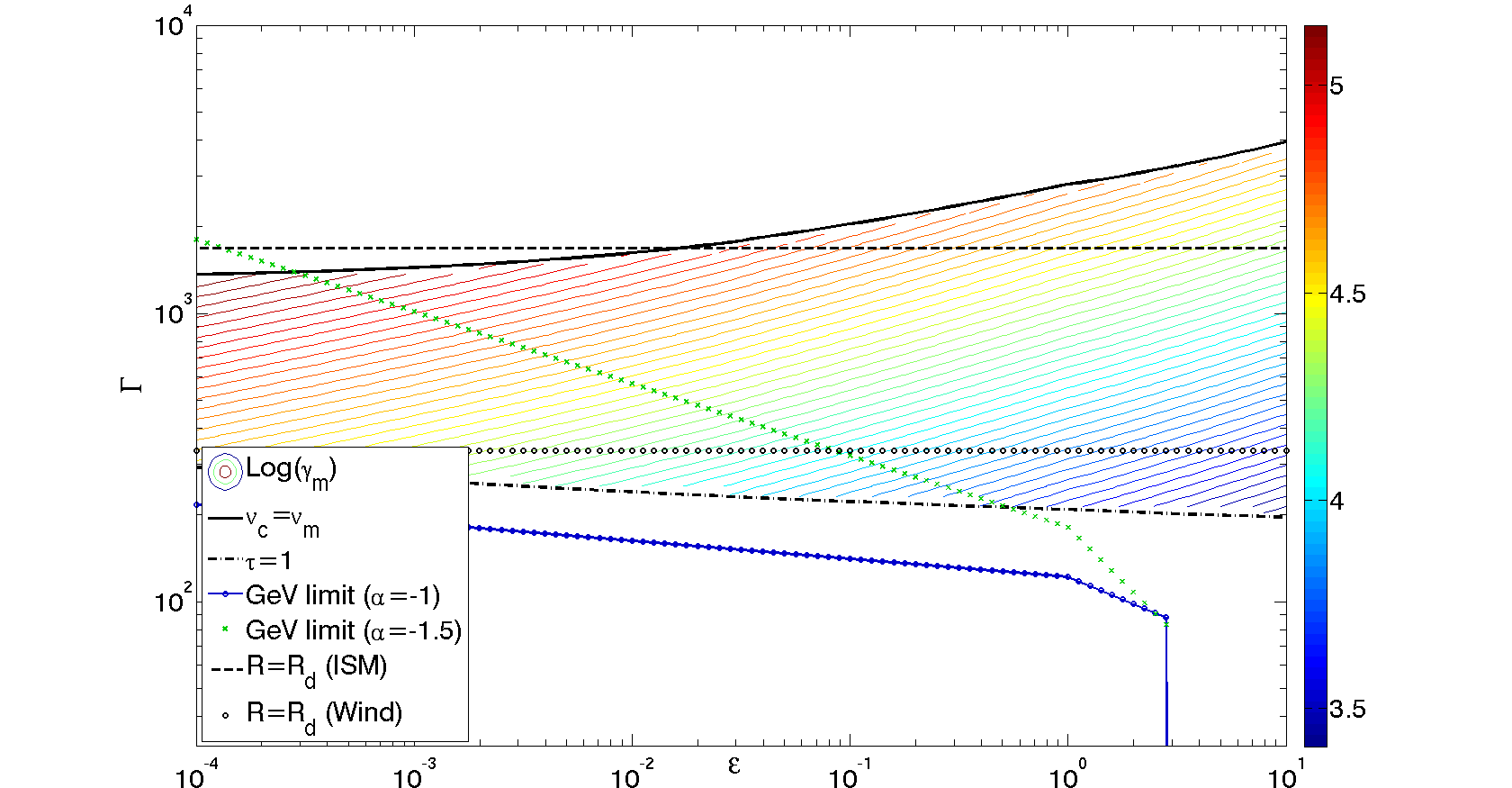}
\caption
{\small The range of GRB parameters for typical observables ($t_{p.3}$,$F_{-26.2}$,$d_{28}$,$\nu_{p,300}$) in
the $(\epsilon, \G)$ plane for $k=1$. Plotted here 
(from top left in clockwise order) are: $R, B, \gamma_m, N_{e}$.
Colours depict the values of these parameters.
The conditions $\tau<1$ (bottom horizontal line) and $\nu_c<\nu_m$ (top curved line) impose strict limits
on the parameter space. The area between them is available for Synchrotron solutions.
The requirement that the SSC signal is below the observational limits in the GeV range, leads to lower limits on $\G(\epsilon)$
which are depicted by the blue circles for the typical lower spectral index ($\alpha=-1$) and by green X's for the expected slope in the fast cooling
regime ($\alpha=-1.5$). The sharp cut-off in these limits corresponds to the transition between $\frac{F_{SSC}}{\nu_mF_{\nu,m}}=Y_0<0.1\eta_{LAT-1}$ where the total
up-scattered flux is below the observational limit, to $\frac{F_{SSC}}{\nu_mF_{\nu,m}}=Y_0>0.1\eta_{LAT-1}$ where the up-scattered signal should peak at high enough energies
so that the flux residing in the observed band is sufficiently low to account for the observations.
These are cases 1 and 2 in Fig. \ref{fig:GeVopt} correspondingly.
Two other conditions $R<R_{dec}(\mbox{Wind})$ and $R<R_{dec}(\mbox{ISM})$ impose softer
limits (the areas below these lines are allowed for each case) on the possible solutions as they both depend on additional unknown parameters ($A_*$ and $n_0$).
}\label{fig:syn_k1}
\end{figure}

\begin{figure} [h]
\centering
\epsscale{1.1}
\plottwo{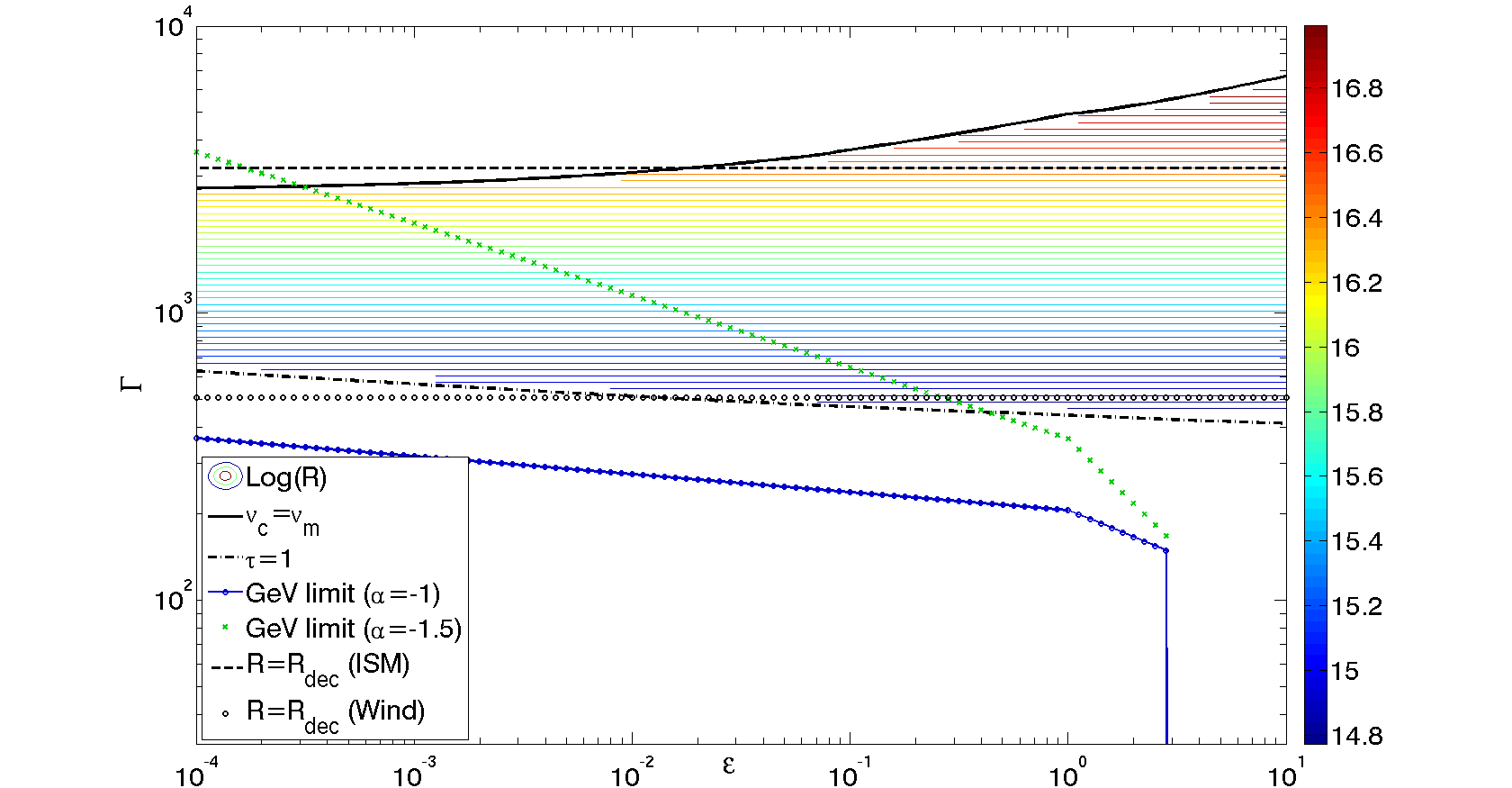}{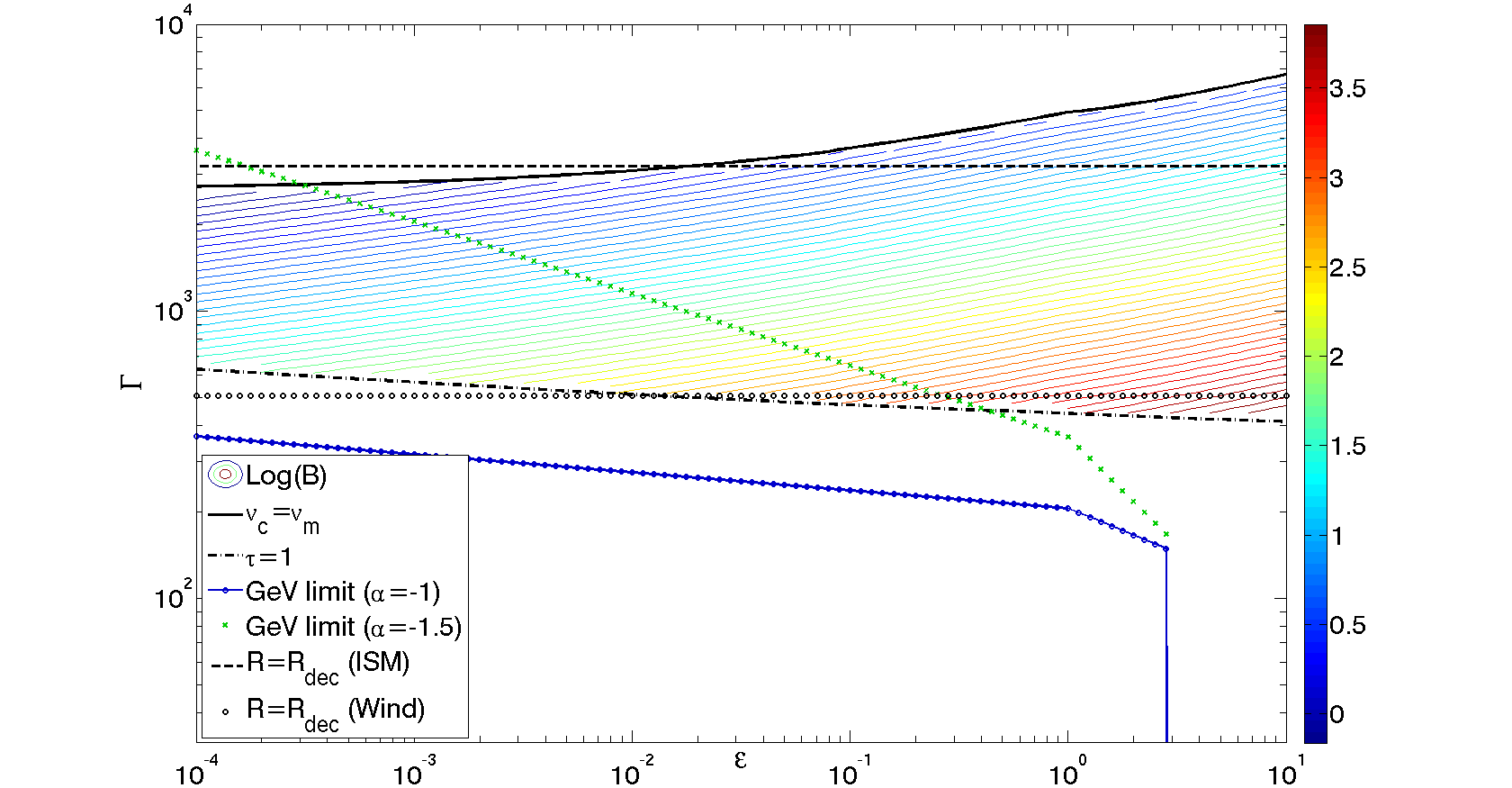}\\
\plottwo{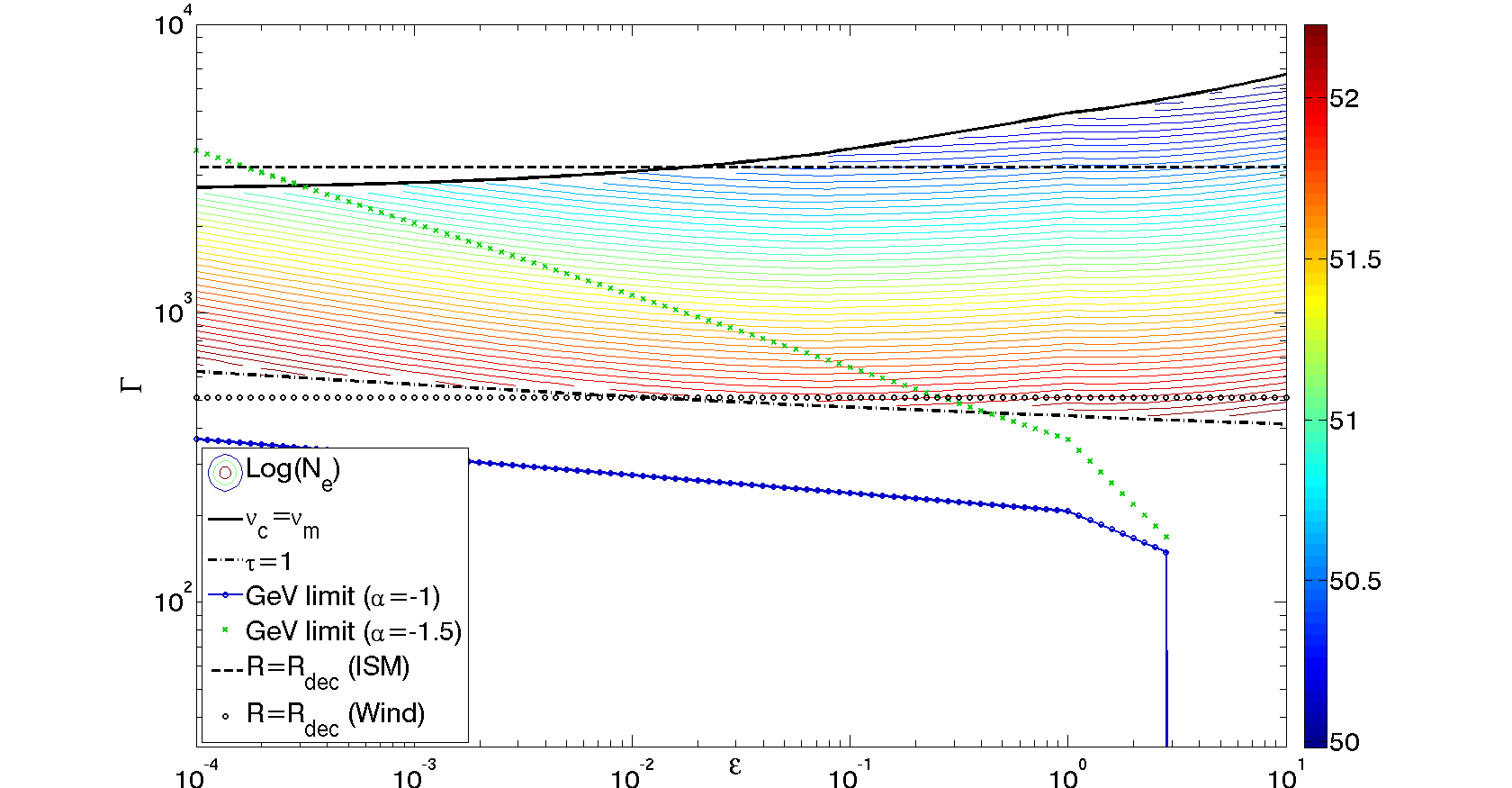}{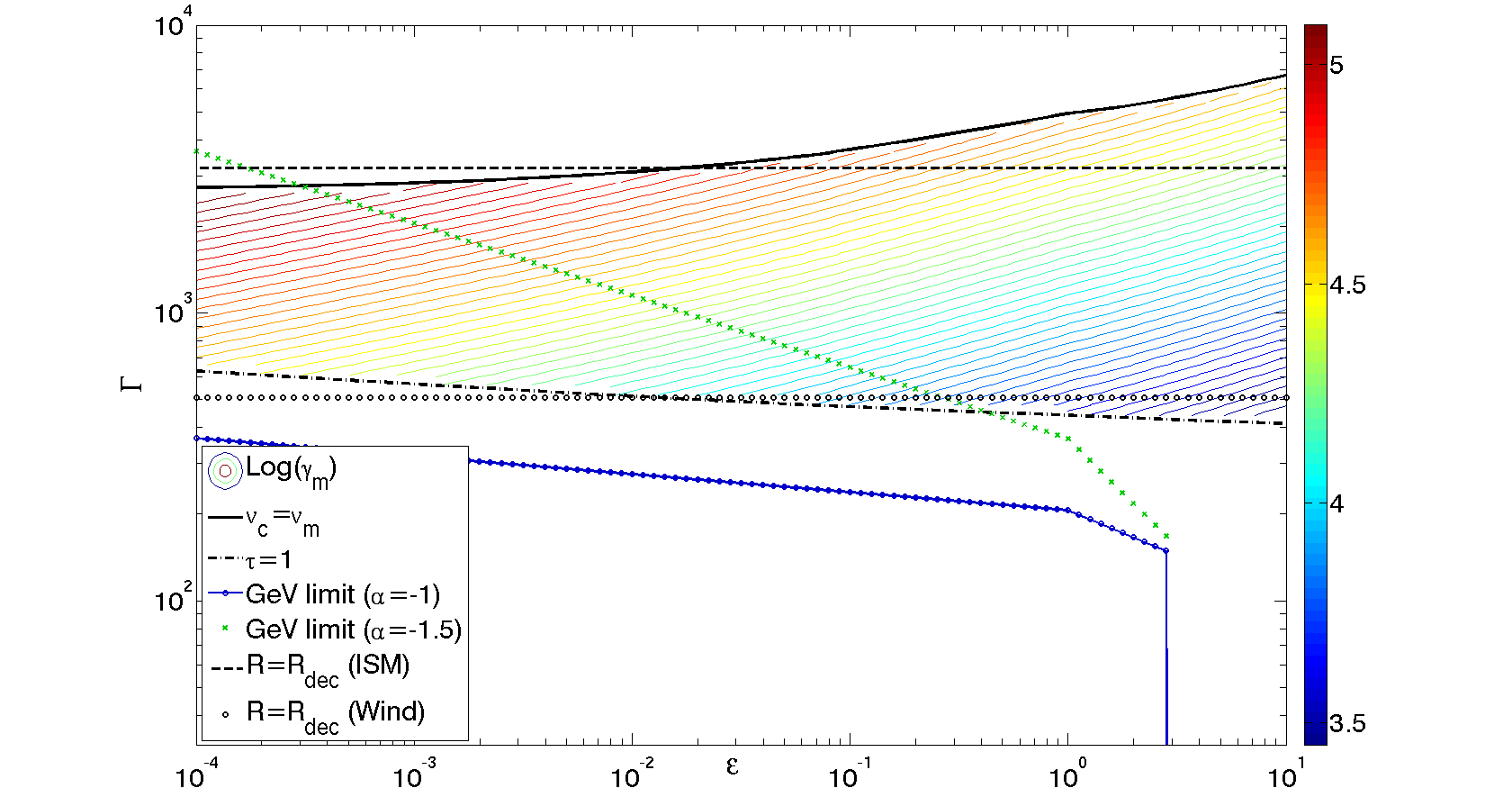}
\caption
{\small Same as Fig. \ref{fig:syn_k1} but for $k=10$.}\label{fig:syn_k10}
\end{figure}

Figs. \ref{fig:syn_k1}, \ref{fig:syn_k10} depict the different GRB parameters superimposed on the
allowed space in the $(\epsilon, \G)$ plane arising from the above conditions.
We plot here results for $k=1$ and $k=10$.
The condition that the SSC flux resides below the observational limits within the frequency range up to the maximal observed photon energy
is plotted for  $\nu_{max}=71$GeV (source frame) corresponding to the highest (prompt) energy photon to date, from 080916C.
They are drawn for two cases of the lower spectral slope: $\alpha=-1$ and $\alpha=-1.5$.
We choose to take the most extreme values for this limit, in order to show that even in this case, there is still reasonable parmeter space
in the $(\epsilon, \G)$ plane. We return to this extreme case in greater detail in \S \ref{sec:paircreation}.

Several characteristic features can be seen in  these figures. First, $R$ is directly related to the pulse duration and bulk Lorentz factor via Eq. \ref{t_p},
and it is independent of $\epsilon$. For the canonical observed parameters we use here, $R$ spans two orders of magnitude,
$10^{15}k^{-1/8}(k+1)^{1/24}t_{p,.3}\mbox{cm}<R<10^{17}k^{-3/8}(k+1)^{1/8}t_{p,.3}\mbox{cm}$, and it is  relatively large.
$B$ is the most fluctuating parameter spanning almost 4 orders of magnitude:
$3k^{-5/16}(k+1)^{1/16}\nu_{p,300}^{1/2}t_{p,.3}^{-1} F_{-26.2}^{1/2} d_{28}\mbox{ Gauss}<B<10^4k^{1/16}(k+1)^{-3/16}\nu_{p,300}^{1/2}t_{p,.3}^{-1} F_{-26.2}^{1/2} d_{28}\mbox{ Gauss}$.
The lines of constant $B$ are almost parallel to the $\nu_m=\nu_c$ line, which means that the value of $B$ is
almost a direct representative of $x_{ins}$, the fraction of instantaneously emitting electrons relative to the total number of electrons.
The allowed range for $\gamma_m$ is:
$3\times 10^3k^{3/16}(k+1)^{-11/48}\nu_{p,300}^{1/4} t_{p,.3}^{1/2} F_{-26.2}^{-1/4} d_{28}^{-1/2}<\gamma_m<10^5k^{1/16}(k+1)^{-3/16}\nu_{p,300}^{1/4} t_{p,.3}^{1/2} F_{-26.2}^{-1/4} d_{28}^{-1/2}$.
These high values of $\gamma_m$ (compared with  $m_p/m_e$)
are extremely significant for GRB models in which the electrons are heated by shocks and the initial energy resides in protons.
In these cases, it is necessary that only a small fraction of electrons will be heated to relativistic velocities in order
to allow them to reach such high energies \citep{Daigne(1998),Bosnjak(2009)}. We show this explicitly for the internal shocks model in \S \ref{intshocks}.
In addition, we observe that $\nu_{KN}$ is expected to lie in the range:
$250k^{1/8}(k+1)^{7/24}\nu_{p,300}^{1/4} t_{p,.3}^{1/2} F_{-26.2}^{-1/4} d_{28}^{-1/2}\mbox{GeV}<\nu_{KN}<1.7k^{-1/8}(k+1)^{3/8}\nu_{p,300}^{1/4} t_{p,.3}^{1/2} F_{-26.2}^{-1/4} d_{28}^{-1/2}\mbox{TeV}.$
Therefore, even with no knowledge of the LAT observations, we could have expected that the SSC does not peak in the GeV but typically at least two orders
of magnitude above. This is a direct consequence of the large values of $\gamma_m$ required for this solution.
$N_e$, is expected to lie between:
$10^{50}k^{-1/8}(k+1)^{-7/24}\nu_{p,300}^{3/4} t_{p,.3}^{1/2} F_{-26.2}^{5/4} d_{28}^{5/2}<N_e<10^{52}k^{1/8}(k+1)^{-3/8}\nu_{p,300}^{3/4} t_{p,.3}^{1/2} F_{-26.2}^{5/4} d_{28}^{5/2}$.

Even for the highest observed photon energy of 71 GeV, we see that the lower limits on $\Gamma$ that arise from the SSC flux in the GeV are less constraining
than the lower limits that arise from the optical depth. However, the SSC limits become very strong for small values of the lower spectral slope,
as expected in the fast cooling synchrotron regime. For the expected $\alpha=-1.5$ (which is found for $10\%$ of the GRBs in the GBM and BATSE samples)
 $\epsilon\gtrsim3 \times 10^{-3}$ for $k=1$ ($\epsilon\gtrsim2 \times 10^{-2}$ for $k=10$).
In addition, very negative values of $\alpha$, push solutions with relatively low $\epsilon$ towards the
marginally fast solution ($\nu_c\approx\nu_m$) as discussed in \S \ref{specshape}.
By Eq. \ref{tau2}, the lower limit on $\G$ varies as $k^{-3/48}(1+K)^{25/48}$ and therefore increases for $k\lessgtr1$.
The upper limit that arises from $\nu_c<\nu_m$ scales as $k^{-3/16}(1+K)^{9/16}$, and also increases for $k\lessgtr1$.
similarly  the allowed parameter space increases with
\footnote{The less rigid upper limit on the parameter space arising from $R<R_{dec}$ also increases with $k$, either as $(k+1)^{3/8}$ for an ISM
or as $(k+1)^{1/4}$ for a wind environment.} $k\lessgtr1$.

The main effect of increasing $k$ is to increase the allowed values for $\G$.
The allowed range for $\G$ is roughly $150<\G<3000$ for $k=1$ and $300<\G<6000$ for $k=10$.
This range is narrowed down by considering the pair creation opacity limits and it will be addressed again in \S \ref{sec:paircreation}.
Interestingly, increasing $k$ does not change significantly other parameters of the model, $R, B, \gamma_m, \mbox{and} N_{e}$.
$R \propto \Gamma^2/k$ implies that the lower limit on the radius  scales as $k^{-1/8}(k+1)^{1/24}$
while the upper limit scales as $k^{-3/8}(k+1)^{1/8}$. Thus, even for relatively large $k$ one expects
a comparable range of emission radii to the one obtained for $k=1$.
For the magnetic field, the lower limit scales as $k^{-5/16}(k+1)^{1/16}$ 
while the upper limit depends on $k^{1/16}(k+1)^{-3/16}$. Again, the allowed range of $B$ varies very weakly with $k$, and $3$ Gauss and $10^4$ Gauss are strict lower and upper limits.
The range of electrons Lorentz factors varies with a lower limit 
$\propto k^{3/16}(k+1)^{-11/48}$, and an upper limit $\propto k^{1/16}(k+1)^{-3/16}$ yielding a strict upper limit of $\approx10^5$ (obtained for $k=1$) and a weakly varying lower limit of order 3000.
Finally, the lower limit on the number of electrons is $\propto k^{-1/8}(k+1)^{-7/24}$, while the upper limit is $\propto k^{1/8}(k+1)^{-3/8}$
implying a strong upper limit on the number of relativistic electrons $\approx10^{52}$.

\subsection{Energetics and efficiency}
\label{energy_efficiency}
We denote by $E_{int}$ the fraction of the overall energy in the flow, $E_{tot}$, that is dissipated into internal energy.
In turn, part of this energy, $E_{syn}$, is radiated by synchrotron to produce the sub-MeV peak. Thus, we define:
\begin{equation}
E_{syn}=\eta_{syn}E_{int}=\eta_{syn}\eta_{int} E_{tot}\equiv \eta_{tot}E_{tot}. 
\end{equation}
$\eta_{int}$ is the conversion efficiency of bulk to internal energy in the prompt phase, i.e. the
efficiency of the energy dissipation process, and $\eta_{syn}<1$ is the conversion efficiency from internal energy
to radiation. 
GRBs must be highly efficient and must emit a significant fraction of the
total kinetic energy in the sub-MeV to avoid an ``energy crisis'', i.e. $\eta_{tot}$ cannot be much
less than unity \citep{Panaitescu(2002), Granot(2006), Fan(2006)}.

The value of $\eta_{int}$ is determined by the energy dissipation process and is therefore not the subject of this paper.
We remark, however, that in many models $\eta_{int}$ is expected to be relatively low. 
For example, in the internal shocks model it is expected that $\eta_{int}\lesssim0.2$
\citep{Kobayashi(1997), Daigne(1998), Belborodov(2000), Guetta(2001)}.
Notice, however, that an exception is possible,in case the same energy can be reprocessed to internal energy more than once
leading to $\eta_{int}>1$ \citep{Kobayashi(2001)}.
It is unknown how natural this scenario is and we leave the discussion on this possibility for future studies.
We conclude that quite generally $\eta_{syn}$ must be smaller than unity.

\begin{figure}[h]
\centering
\plottwo{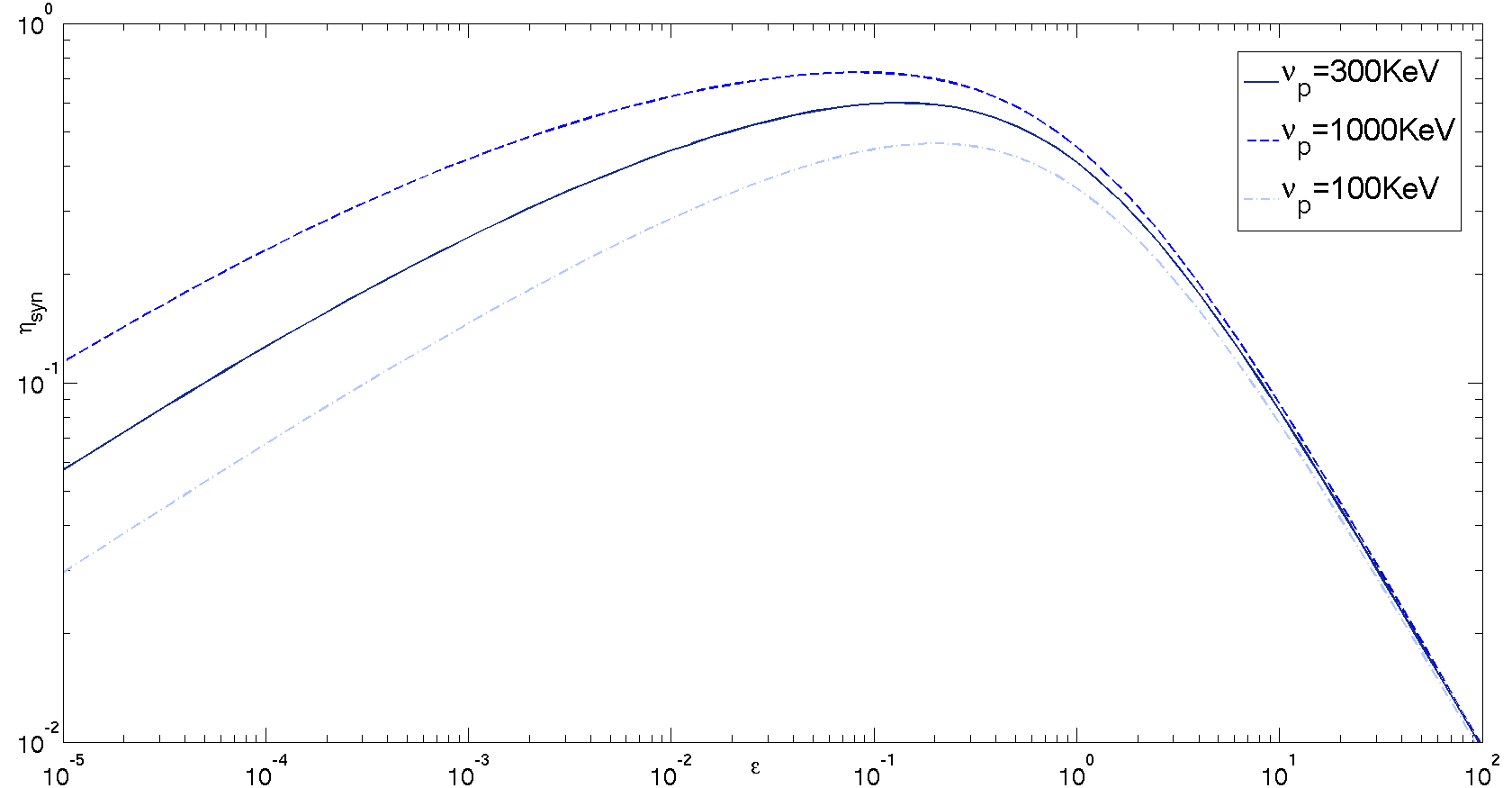}{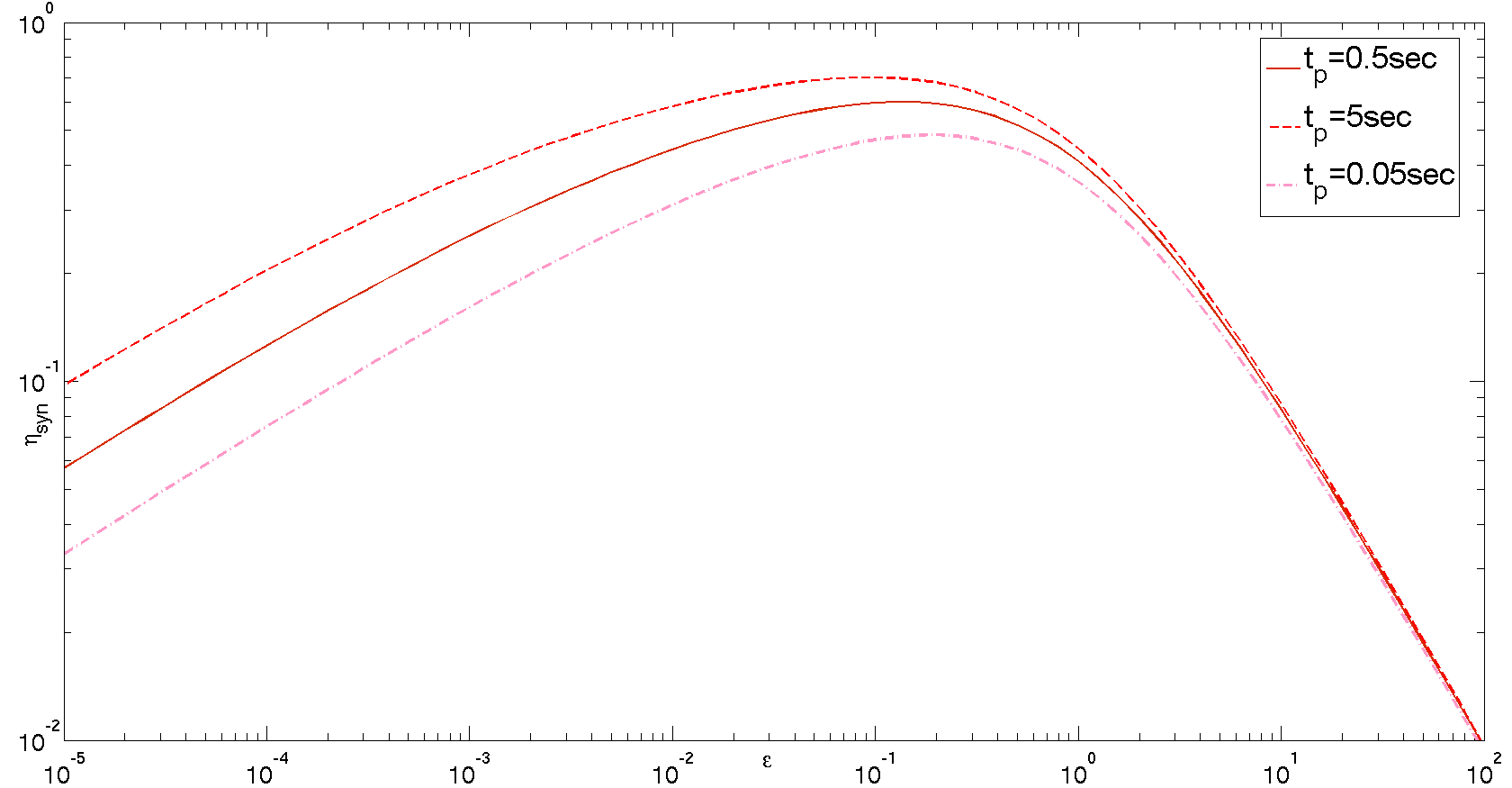}
\caption
{\small An upper limit on the efficiency when the sub-MeV peak is ascosiated with fast cooling synchrotron
emission, as a function of $\epsilon$.
The electrons must carry a significant amount of the energy, but too strong an amount
will lead to a large amount of energy released at frequencies far above the peak by the KN process.
The left panel shows the dependence on $\nu_p$ while the right panel shows the dependence on the
pulse duration.
These plots depict upper limits on the efficiency, as they
do not include the efficiency of the energy dissipation process or energy losses in components
other than magnetic fields or electrons.} \label{fig:eff}
\end{figure}

Two efficiency criteria determine the efficiency of the synchrotron signal.
First, the energy radiated by synchrotron is limited by the energy of the relativistic electrons.
Therefore, we focus here on fast cooling to avoid an extra loss in the efficiency.
This means that $\eta_{syn}\leqslant \epsilon_e$ and therefore $\epsilon_e$ cannot be too small.
Accounting for SSC losses in addition to synchrotron losses, this translates to an
upper limit on $\epsilon$.
On the other hand, $\epsilon$ is also bounded from below by the
requirement that SSC loses are not too large.
$\frac{F_{SSC}}{\nu_m F_{\nu,m}}\equiv Y_0$ is a function of $\epsilon$,
independent of the bulk Lorentz factor, but in the KN regime it may still depend on $k$ and
on the various observed parameters. As seen from Eq. \ref{Y0}, apart for the peak frequency and pulse duration
(which can span over a large range of values),
all of these dependencies are very weak and have no significant effect on the overall efficiency.
We examine the dependence of $\eta_{syn}$ on $\nu_p$ and $t_p$ in Fig. \ref{fig:eff}. 

Combining all these efficiency factors, we get the overall efficiency:
\begin{equation}
\label{etatot}
\eta_{tot}=\eta_{int} \eta_{syn}=\frac{\eta_{int} \epsilon_e}{1+Y_0}.
\end{equation}
As $\eta_{int}$ alone is already expected to be relatively low, this limits
$\eta_{syn}\gtrsim0.1$ in order to assure that the overall efficiency satisfies: 
$\eta_{tot}>10^{-2}$. Thus, we must have: $10^{-4}<\epsilon<10$ (see Fig. \ref{fig:eff}).
Notice that this result is independent of the bulk Lorentz factor, and it is weakly dependent on $k$ and the
observed parameters. Out of these, the strongest dependence is on $\nu_p$.
$\nu_p\approx100$KeV ($\nu_p\approx1000$KeV) increases (decreases) the lower limit on $\epsilon$
to about $10^{-3}$ ($10^{-5}$). Although low $\nu_{peak}$ results in low efficiencies,
such bursts are usually observed with lower luminosities, and the energetic requirements for
such bursts are not unreasonable.
We return to this point in more detail in \S \ref{Epeak}.
The efficiency depends more weakly on the pulse duration, but the latter
spans at least three orders of magnitude and thus can cause noticeable change. Taking $t_p\approx5\times10^{-2}$sec ($t_p\approx5$sec)
yields $\epsilon\gtrsim3\times 10^{-4}$ ($\epsilon\gtrsim10^{-5}$).
Later on, we find that high magnetization is favored by other reasons as well.

\subsection{The spectral shape}
\label{specshape}
The solutions given in \ref{synchsol} consider only the peak frequency and the peak flux, disregarding the spectral shape.
In the fast cooling scenario (considered here) the
slope of the photons' number spectrum above the peak of $\nu F_{\nu}$ is 
($\nu^\beta\propto \nu^{-(p+2)/2}$) \citep{Sari(1997)}. This is consistent with the average observed slope ($\beta=-2.3$)
for $p\approx2.6$ which is approximately the expected value for $p$ in case the electrons are accelerated by the Fermi mechanism \citep{Achterberg(2001)}
(Although it is not clear that the spread of $\beta$ is consistent with the expected spread in $p$ \citep{Shen(2006),Starling(2008),Curran(2010)}).
However, the synchrotron fast cooling slope below the peak of $\nu F_{\nu}$
($\nu^\alpha\propto\nu^{-1.5}$) \citep{Cohen(1997), Sari(1998), Ghisellini(2000)}, is inconsistent with the average observed slope $\alpha=-1$ 
and $90\%$ of bursts have steeper slopes than this \citep{Preece(2000), Ghirlanda(2002), Kaneko(2006), Nava(2011)}. 
Furthermore, the lower energy slope in about $40\%$ of GRBs is $\alpha>-2/3$ \citep{Nava(2011)} which is impossible in
a synchrotron scenario, even in the case of slow cooling. This is known as the synchrotron ``line of death'' \citep{Preece(1998)}.
An extra concern, is that the break observed at $\nu_{peak}$ is often too sharp to be compatible with the smooth transition
expected from the synchrotron model \citep{Pelaez(1994)}.

\begin{figure}[h]
\centering
\plottwo{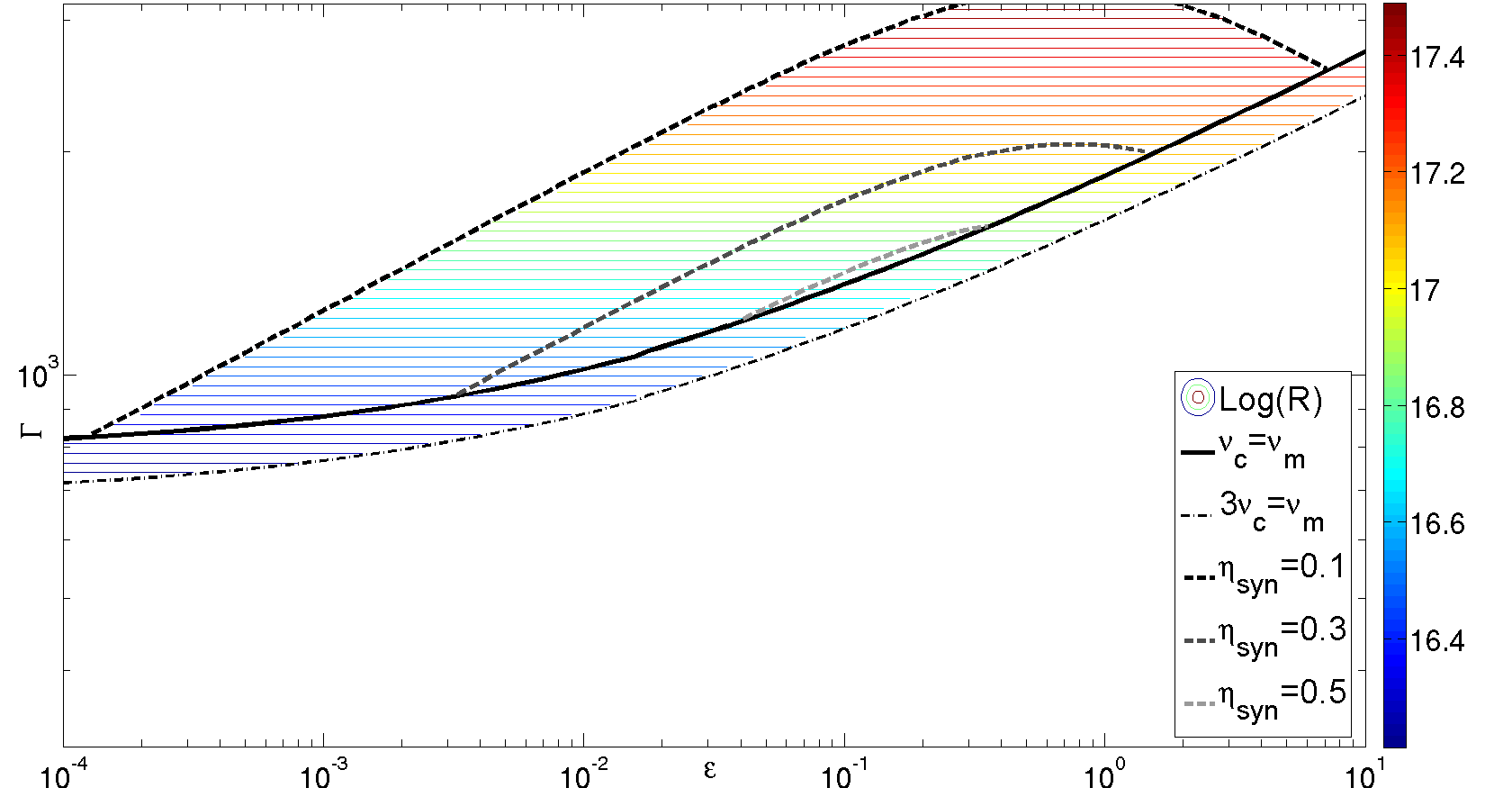}{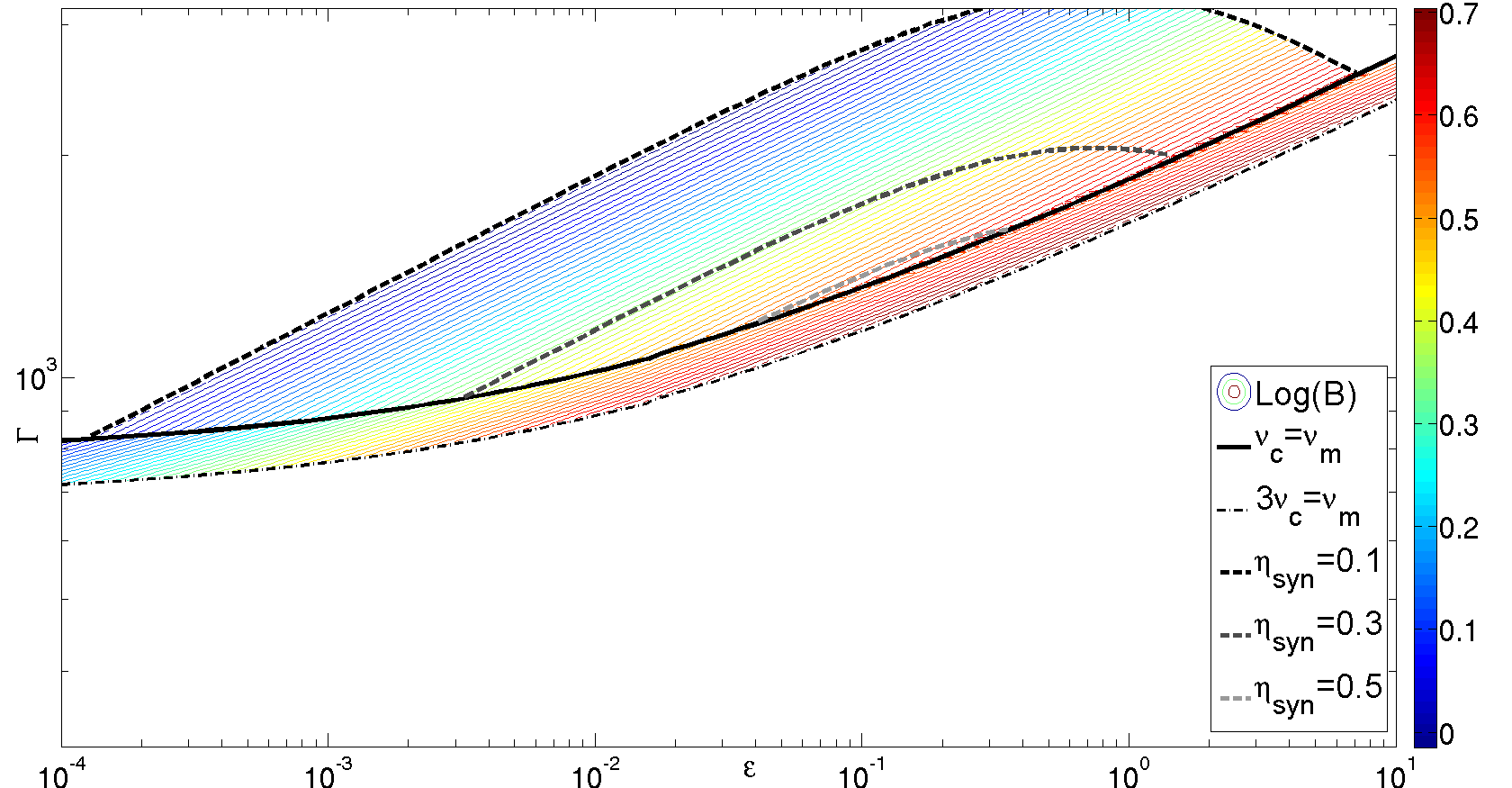}\\
\plottwo{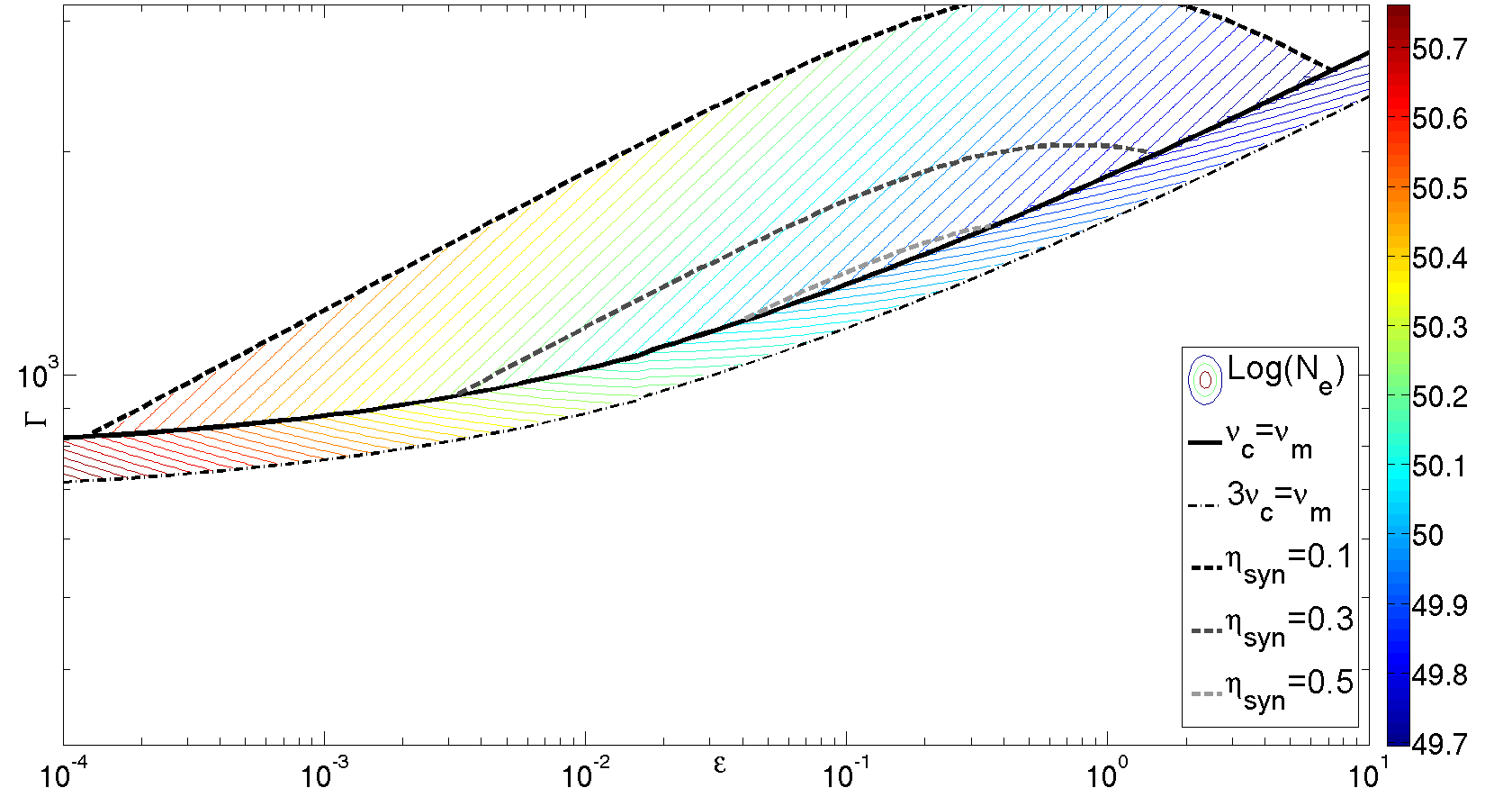}{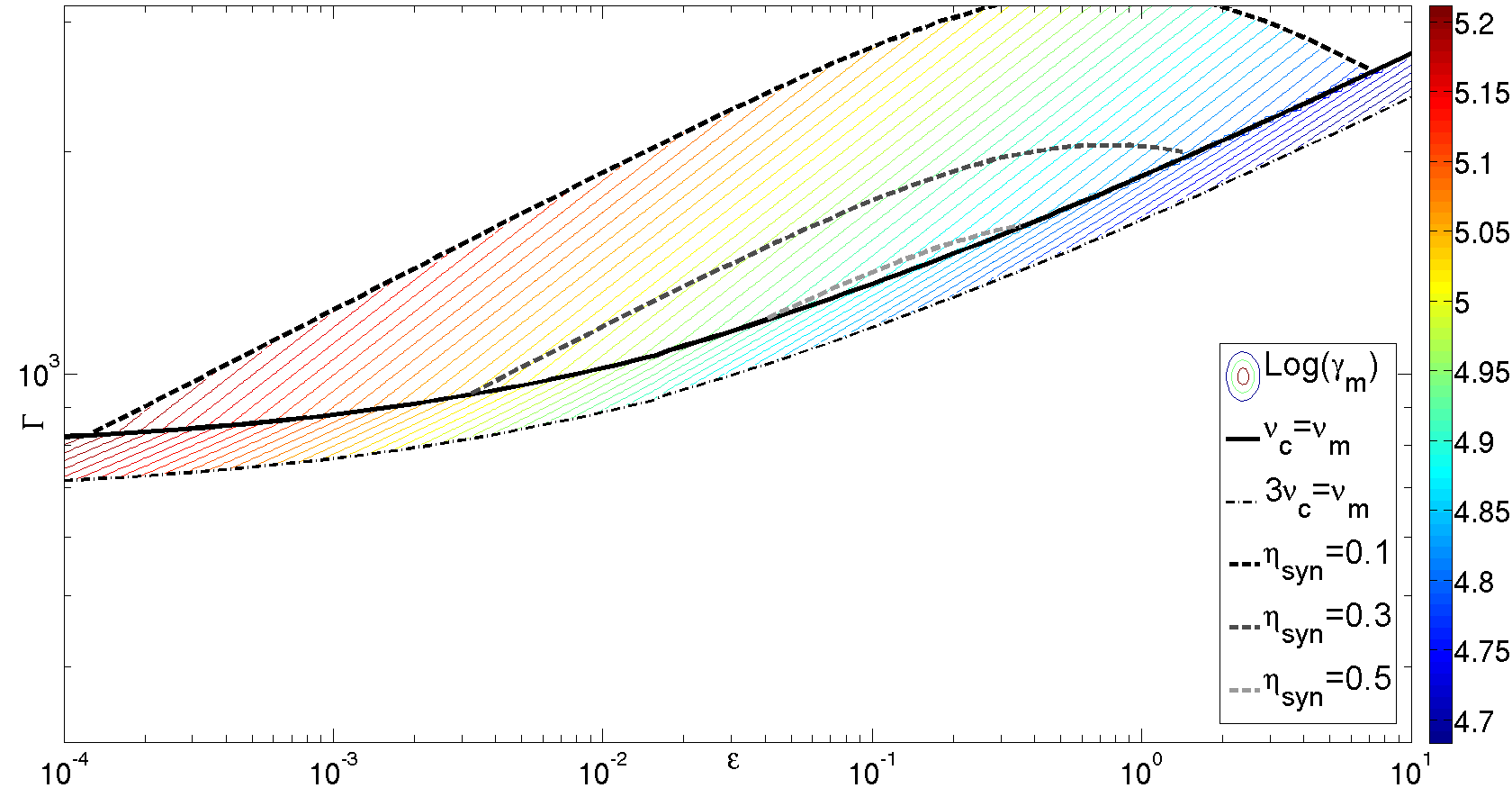}
\caption
{\small The range of GRB parameters for a marginally fast solution in
the $(\epsilon, \G)$ plane for $k=1$, $\xi=1$. Plotted here 
(from top left in clockwise order) are: $R, B, \gamma_m, N_e$.
Below the dashed line ($\nu_m=3\nu_c$) there is a sufficiently large frequency range between $\nu_c$ and $\nu_m$
leading to $\alpha=-1.5$ below the sub-MeV peak which is incompatible with the marginally fast solution.
Above the black dot-dashed line, the radiative efficiency falls below 0.1.
In addition are plotted the lines of $\eta_{syn}=0.5, 0.3$ in grey dot-dashed lines.}\label{fig:marginal}
\end{figure}

The synchrotron ``fast cooling - line of death'' problem, the more serious of the two, can be partially resolved if the electrons are marginally fast cooling \citep{Daigne(2011)} with $\nu_m \approx \nu_c$.
This requires, of course, an additional fine tuning mechanism that keeps this condition satisfied.
To examine the possible parameter phase space with a ``marginal fast cooling" we define a marginally fast solution as one that obeys one of two criteria.
Either it is a fast cooling solution ($\nu_m>\nu_c$) that obeys: $\nu_m<3\nu_c$ (If $\nu_m$ were much larger, there would be a large range of frequencies where $\alpha=-1.5$ and the
``fast cooling - line of death'' problem would not be solved);
or it is a a slow cooling solution ($\nu_c>\nu_m$) (where $\alpha=-2/3$ is always valid so long as we associate $\nu_m$, and not $\nu_c$, with the observed sub-MeV peak) which is sufficiently
efficient, with: $\eta_{syn}>0.1$.
This family of solutions, is seen in Fig. \ref{fig:marginal} and it occupies a significant area in the parameter space.
The break in the slopes of the parameters in the ($\G, \epsilon$) plane is due to the switch from the fast to slow cooling regime. 
The upper curve is the limit on the marginally fast solution due to efficiency. These lines cross the $\nu_c=\nu_m$ line at $\epsilon\approx10^{-4}$ and $\epsilon\approx10$
which is where the fast cooling efficiency falls below 0.1 (see Fig. \ref{fig:eff}).
The marginally-fast solutions are characterized by $t_c(\gamma_m) \approx t_p$, and therefore $x_{int}\approx1$ and the instantaneous number of emitting electrons is about the same as the total number of relativistic electrons.
The allowed region has high values of $\G$ and $\gamma_m$: $\G\gtrsim700$, $\gamma_m \approx 10^5$, relatively small values of $N_e$: $N_e\approx10^{50}$,
a weak (and almost constant) magnetic field: $B\approx 10 \mbox{ Gauss}$
and a large emission radius $R\gtrsim3 \times 10^{16}cm$. A canonical wind solution is ruled out for these solutions.
Another partial solution to the ``line of death'' problem, can be achieved in case SSC occurs in the KN regime, as was shown in \S \ref{sec:SSC} to
be the generic case. In such cases, detailed modeling, shows that $\alpha$ can be increased from -1.5 to -1 \citep{Derishev(2003),Nakar(2009),Wang(2009),Daigne(2011),BD(2012)}.
However, this solution requires very low values of $\epsilon$ and  with $\epsilon\gtrsim10^{-4}$, as required by efficiency considerations (see \S \ref{energy_efficiency}),
one is still limited to $\alpha<-1.2$.

A complete solution to the ``line of death'' problem within synchrotron emission can be achieved if the
synchrotron self absorption frequency, $\nu_{SSA}$, is
sufficiently close to $\nu_m$.
Extrapolating the fast cooling synchrotron flux below $\nu_m$ and equating
it to the black body limit \citep{Granot(2000)} we can estimate $\nu_{SSA}$:
\begin{equation}
 \label{SSA}
\frac{2\nu_{SSA}^2}{c^2}\gamma(\nu_{SSA}) \G m_e c^2 \frac{R^2}{4\G^2d_L^2}=F _{\nu ,max} (\nu_{SSA}/\nu_m)^{-1/2}
\end{equation}
where $\gamma(\nu_{SSA})$ is the typical thermal Lorentz factor of electrons radiating at $\nu_{SSA}$.
Eq. \ref{SSA} yields a (source frame) self absorption frequency of:
\begin{equation}
 \label{SSA2}
\frac{\nu_{SSA}}{\nu_{p,300}}=9.2\times10^{-5}(1+Y_0)^{1/12}\G_2^{-4/3} (\epsilon)^{1/12} k^{-1/12}(k+1)^{11/12} \nu_{p,300}^{-3/4}t_{p,.3}^{-5/6} F_{-26.2}^{5/12} d_{28}^{5/6}.
\end{equation}
Typically, $\nu_{SSA}$ is about 4 orders of magnitude bellow $\nu_p$ and it
depends weakly on the various parameters.
It is therefore, unlikely that self absorption has any significant effect on $\alpha$.

A final possibility that should be mentioned in the ``line of death'' context is that $\alpha$ may be affected by the presence of a sub-dominant
thermal component which is expected to occur in the fireball model at a level of a few percent of the total prompt energy \citep{Meszaros(2000),Daigne(2002),Nakar(2005)}.
Detailed spectral modeling, shows that adding a weak sub-dominant thermal component usually causes $\alpha$ to decrease \citep{Guiriec(2011),Guiriec(2012)}.

\subsection{The Distribution of $E_{peak} \backslash L_{peak}$}
\label{Epeak}

There have been many claims that the distribution of $E_{peak}$ is rather narrow
\citep{Band(1993), Mallozzi(1995), Brainerd(1999), Schaefer(2003)} .
Typically $E_{peak}$ values are in the range $150\mbox{ KeV}<E_{peak}<700$ KeV \citep{Kaneko(2006)}.
It is not obvious that this observation reflects the intrinsic properties of GRBs and is not caused by some selection effect \citep{Shahmoradi(2010)}. Softer bursts have been detected by Beppo-SAX and HETE2,
and  Fermi is finding much higher $E_{peak}$ bursts. 
However, if true, this result is surprising in the general framework of the Synchrotron model, where one expects $E_{peak}=h\nu_m \propto \G \gamma_m^2 B$
and it is not obvious why this specific combination of parameters will remain fairly constant between different pulses and different bursts.
Furthermore, there have been claims that $E_{peak}$ may be correlated with the peak luminosity, $L_{peak}$, \citep{Yonetoku(2010)} and that therefore the distributions of both are not statistically independent.

\begin{figure}[h]
\centering
\plottwo{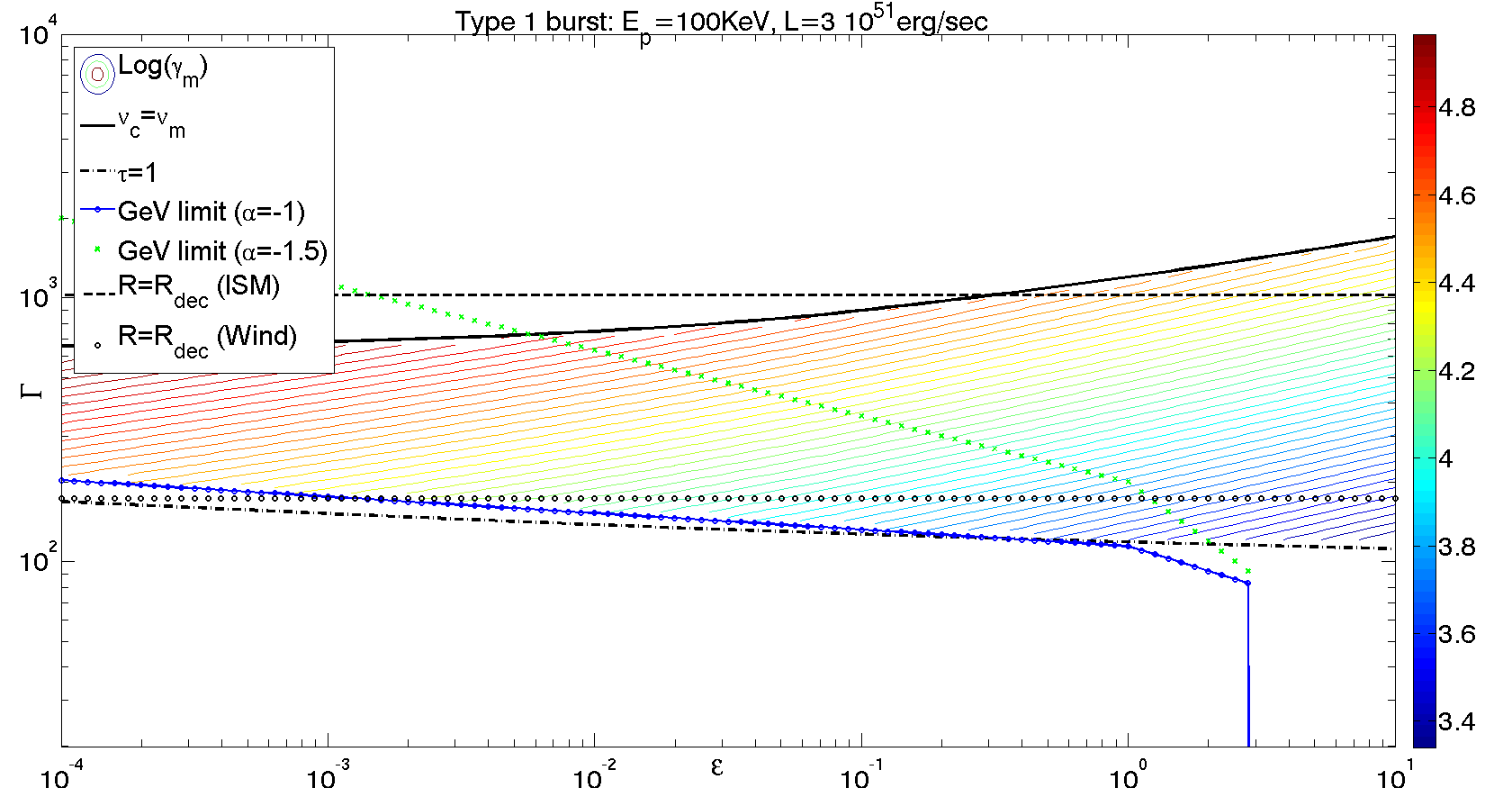}{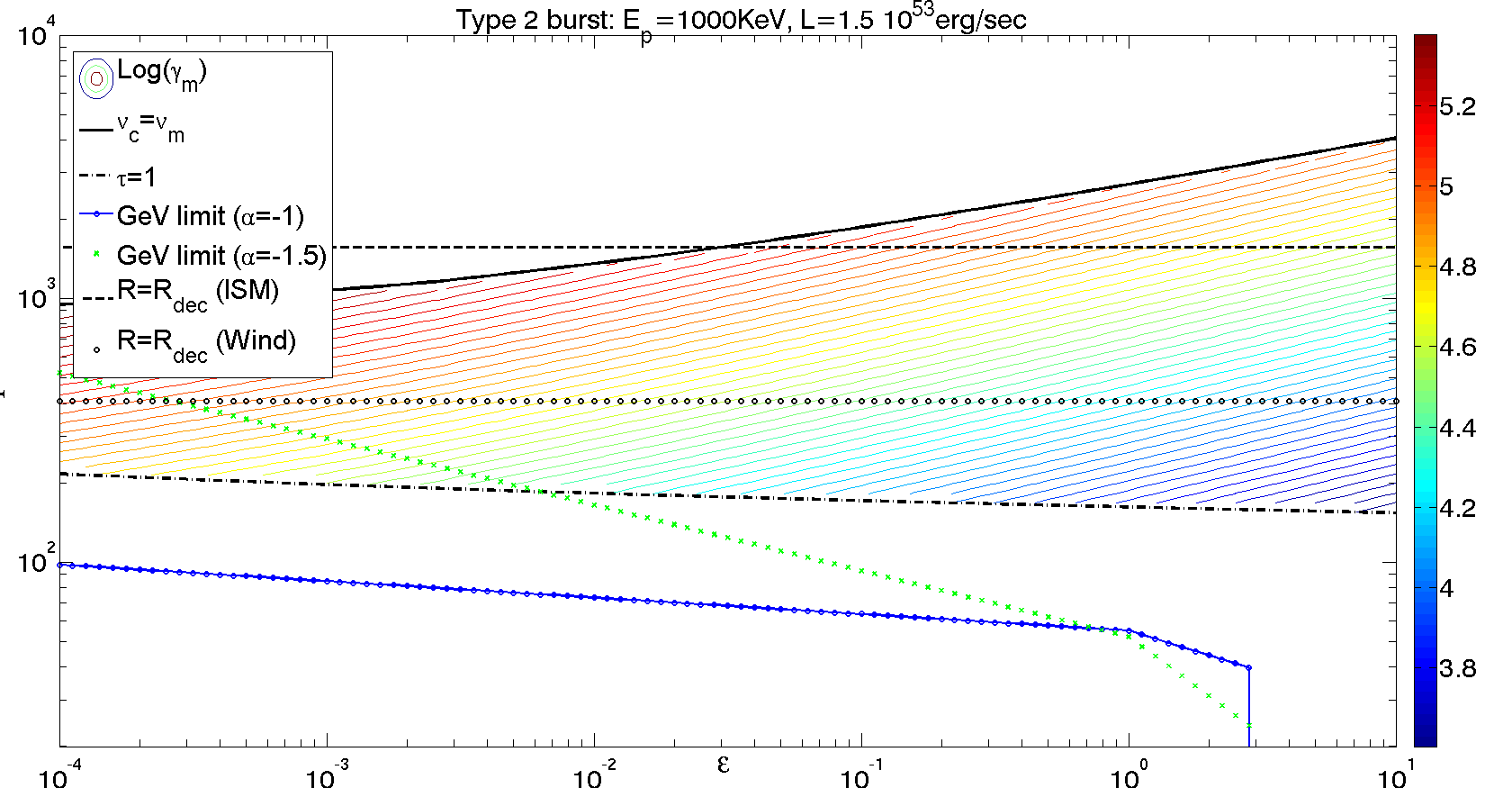}\\
\plottwo{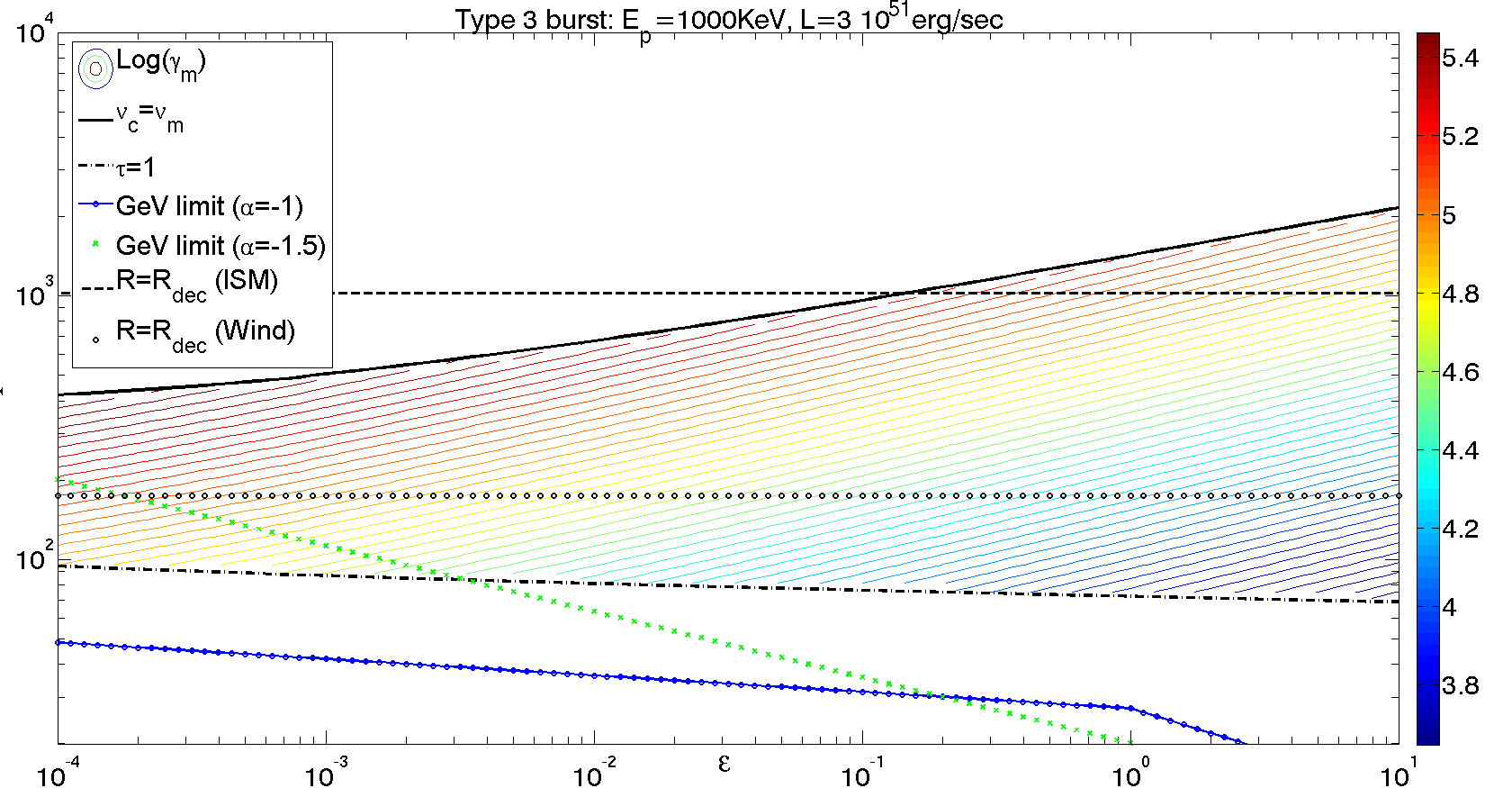}{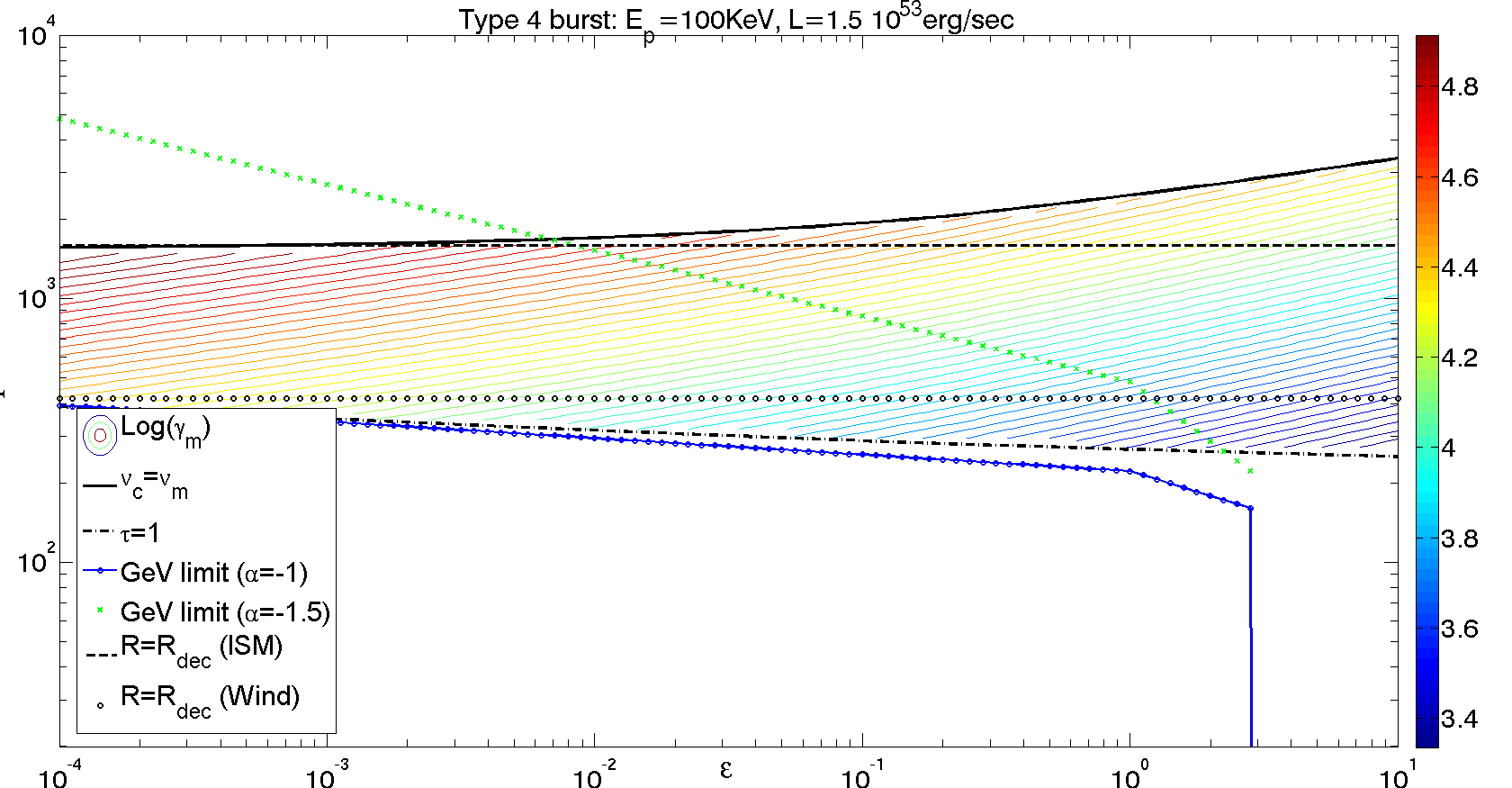}
\caption
{\small The range of $\gamma_m$ (shown by different colours) for four types of bursts
in the $(\epsilon, \G)$ plane for $k=1$, $\xi=1$.
The conditions $\tau<1$ (bottom horizontal line) and $\nu_c<\nu_m$ (top curved line) impose strict limits
on the parameter space. The area between them is available for Synchrotron solutions.
The requirement that the SSC signal is below the observational limits in the GeV range, leads to lower limits on $\G(\epsilon)$
which are depicted by the blue circles for the typical lower spectral index ($\alpha=-1$) and by green X's for the expected slope in the fast cooling
regime ($\alpha=-1.5$).
Two other conditions $R<R_{dec}(\mbox{Wind})$ and $R<R_{dec}(\mbox{ISM})$ impose softer
limits (the areas below these lines are allowed for each case) on the possible solutions.
}\label{fig:fourB}
\end{figure}

\begin{figure} [h]
\centerline{ \plotone{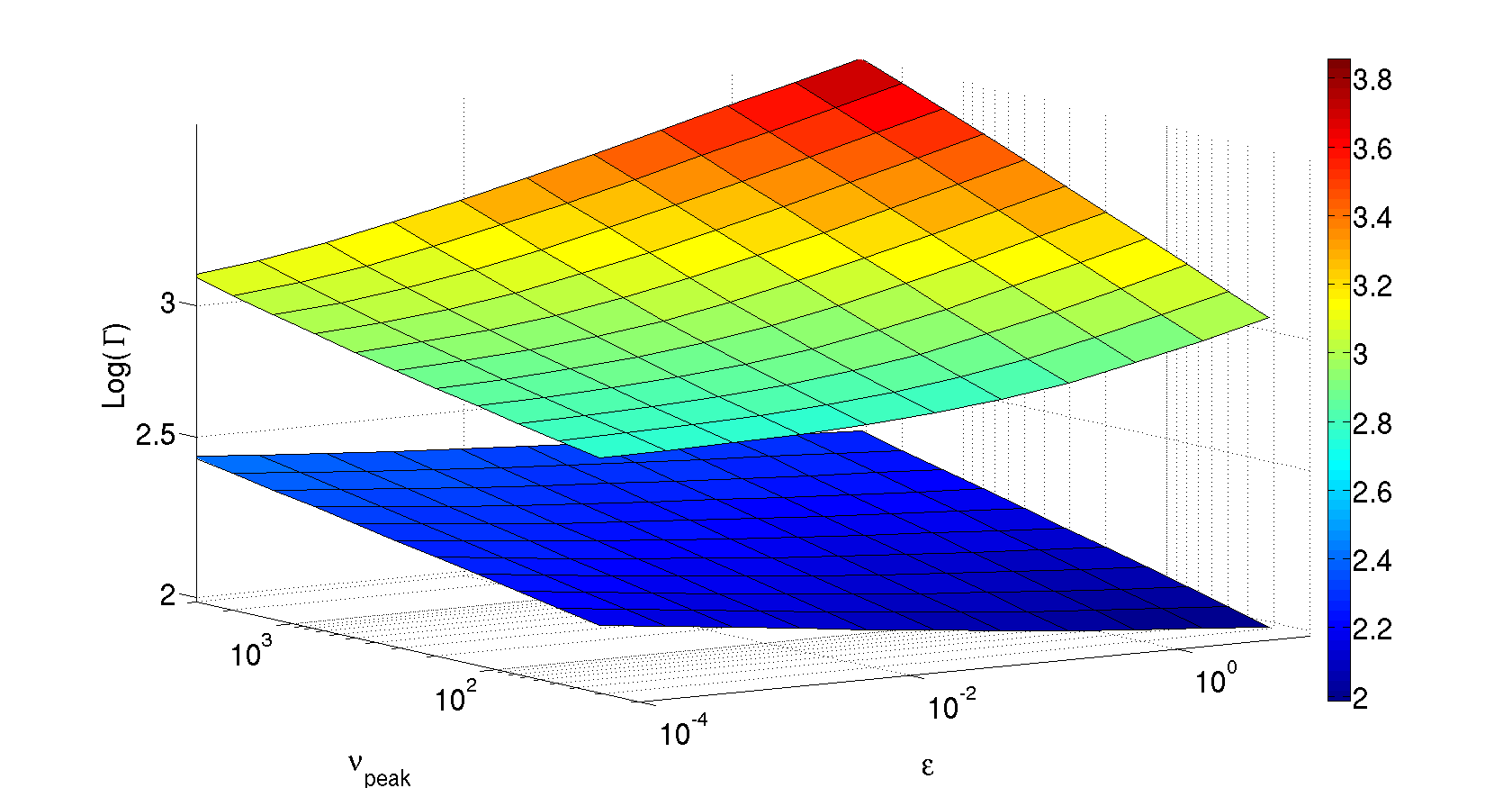}} \caption
{\small Limits on the allowed parameter space due to $\nu_m\geqslant\nu_c$ and $\tau\leq1$ in the ($\epsilon$, $\nu_{peak}$,$\G$) space assuming that
$\nu_{peak}$ and $L_{peak}$ follow the Yonetoku relation \citep{Yonetoku(2010)}.
Fast cooling optically thin synchrotron solution an arise in
the area between the two surfaces. } \label{fig:G3D}
\end{figure}

To examine this issue we examine four characteristic bursts in the $(E_{peak}, L_{peak}$) plane:
\begin{enumerate}
 \item Low $E_{peak}$ ($100KeV$) and low $L_{peak}$ ($3 \times 10^{51}erg/sec$) (low end of Yonetou relation).
 \item High $E_{peak}$ ($1000KeV$) and high $L_{peak}$ ($1.5 \times 10^{53}erg/sec$) (high end of Yonetoku relation). 
 \item High $E_{peak}$ ($1000KeV$) and low $L_{peak}$ ($3 \times 10^{51}erg/sec$) (``above'' the Yonetoku relation).
      None of the bursts with known redshift belong to this category, but this may be an observational effect \citep{NakarPiran(2005),Band(2005)}.
 \item Low $E_{peak}$ ($100KeV$) and high $L_{peak}$ ($1.5 \times 10^{53}erg/sec$) (``below'' the Yonetoku relation).
Such bursts would be relatively easier to detect (many photons) but are not seen in observations. It is therefore believed that they do not exist. 
\end{enumerate}
We are interested in the range of parameters ascosiated with synchrotron solutions
for these bursts. We note that $L_{peak}\propto F_{\nu_{peak}} \nu_{peak} d^2$, so that there is generally
more than one way to realize a specific value of $L_{peak}$ with the different observed parameters. In the following discussion, we
choose to take $d$ to be constant and vary $F_{\nu_{peak}}$. 
In terms of the allowed ranges for the synchrotron parameters we do not find any significant difference between the four typical bursts 
(see Fig. \ref{fig:fourB} for an example of $\gamma_m$ in each of these bursts).
The main difference is a slight change in the limits of the allowed region corresponding to $\nu_c=\nu_m$ and $\tau=1$.
This can also be seen in Fig. \ref{fig:G3D} where we plot those limits in the ($\epsilon$, $\nu_{peak}$) plane assuming that
$\nu_{peak}$ and $L_{peak}$ follow the Yonetoku relation.

\begin{figure} [h]
\centering
\plottwo{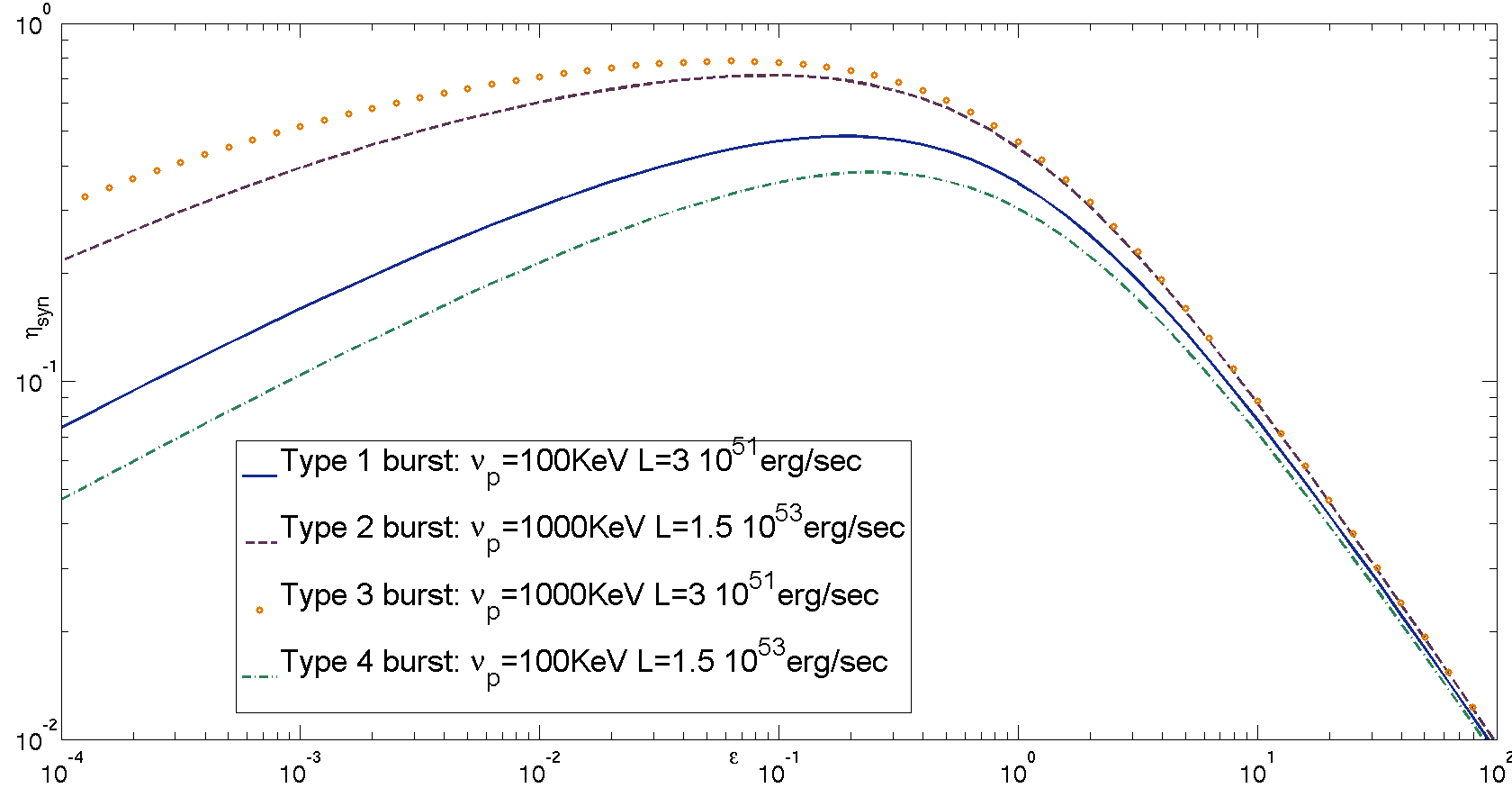}{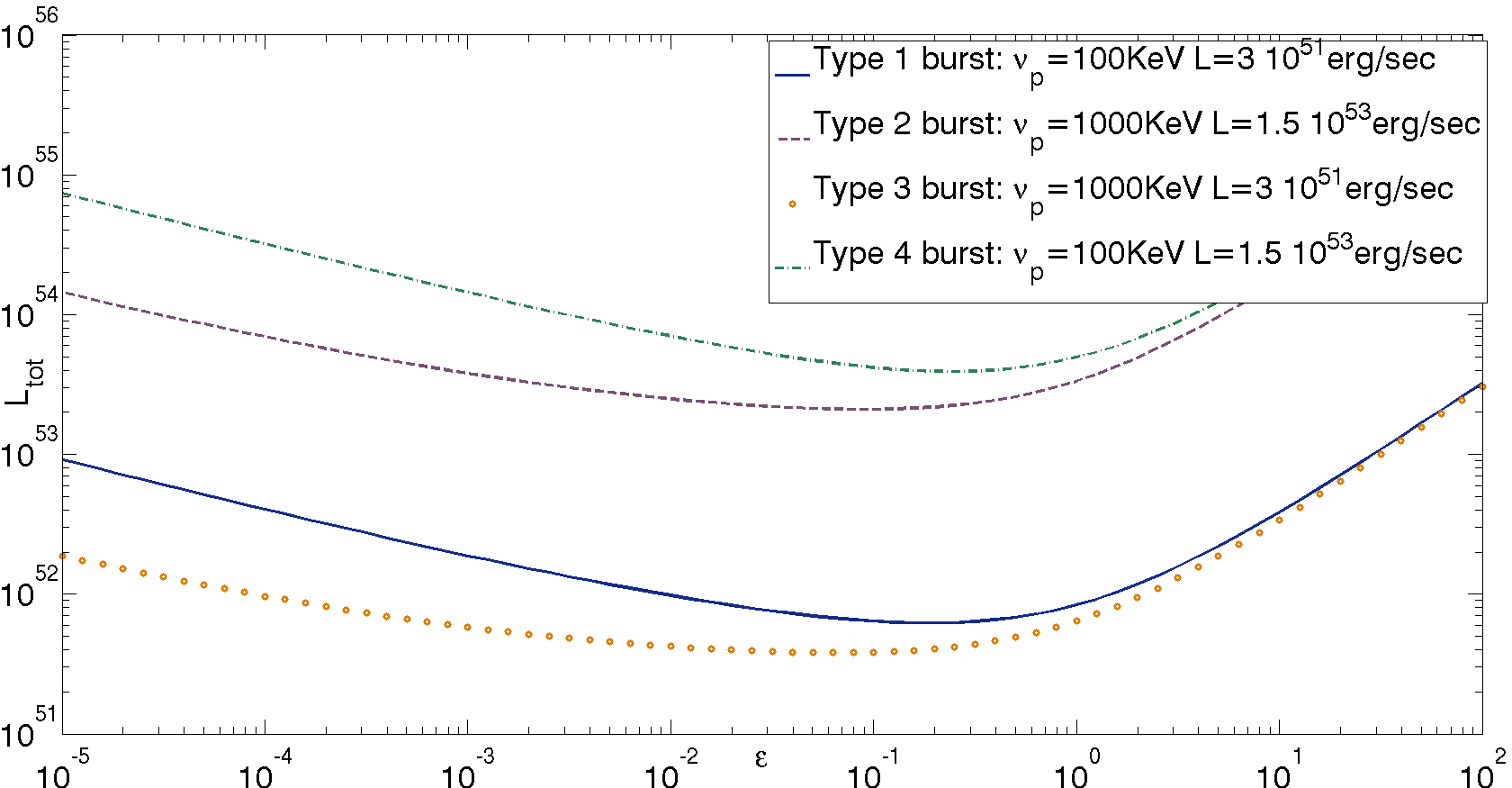}
\caption
{\small The maximal radiative efficiency of the four different types of bursts.
Soft, high luminosity bursts are the least efficient and require huge 
energies at the source. This may explain the lack of such bursts.} \label{fig:foureff}
\end{figure}

In Fig. \ref{fig:foureff} we plot the radiative efficiency and minimal luminosity requirements of these bursts (as described in \S \ref{energy_efficiency}).
A type-4 burst is the most inefficient and has the largest intrinsic power.
Its synchrotron efficiency peaks at less than 0.4, and it could very
likely be less, depending on the value of $\epsilon$, while  its intrinsic power is at least $4\times 10^{53}erg/sec$.
This may pose an explanation as to why practically no soft - high luminosity bursts are seen (i.e.
bursts below the Yonetoku relation). The energetic demands for such bursts are simply too high.
The other class of Yonetoku relation violators, are type 3 bursts. For these, efficiency is high and they cannot be ruled out by the same argument.
However such bursts (hard and weak) have only a small number of photons \citep{NakarPiran(2005),Band(2005)} and are observationally hard to detect.
This might cause fewer bursts of this kind to have a detectable redshift, leading to a natural selection effect.

Unless, typical bursts have very low $\epsilon$, the observed luminosity is much more important than $\nu_{peak}$ in order to
determine the intrinsic power. Therefore, it is possible to produce bursts with low $L_{peak}$ and $\nu_{peak}$ and we conclude that
this model does not naturally predict the claimed narrow distribution of $\nu_{peak}$ by
efficiency arguments alone. As mentioned above, this may actually be considered as an advantage of the synchrotron model, as
it is quite likely that the true peak distribution is not as narrow as previously thought. In this case, the argument would be reversed.
Synchrotron would be a very natural way to produce such large peak dispersion.

\section{A synchrotron solution within the internal shocks model}
\label{intshocks}

We divert here from the main theme of the paper, in which we consider a ``model independent
synchrotron emission" and examine specific additional constraints that arise when considering 
a situation when the energy is stored in the bulk kinetic energy of the flow. This is the case in the 
common internal shocks model, but is also the case in a Poynting flux dominated model if the
magnetic energy has been converted to kinetic energy before the radiation process begun.

The kinetic energy of the initial flow is dominated by protons
\footnote{In principle, many pairs could be created after the shocks collide and the internal energy is released.
This could cause the combined shell to be dominated by pairs, but has no effect on the following consideration which is related to
the initial composition.}.
Denoting the number of protons by $N_p$, we have: $N_p=N_{tot}$ ($2N_{e+}=N_{tot}$ for a pair dominated plasma).
The total (bulk+internal) energy needed to produce a combined shell traveling with a bulk Lorentz factor $\G$, in this scenario, is:
\begin{equation}
\label{Etot}
E_{tot}\approx g \G N_{tot}m_p c^2 (\times \frac{m_e}{m_p} \mbox{for pairs})
\end{equation}
where the factor of $g \approx few$
arises from the fact that the colliding shells must initially have different bulk Lorentz factors in order for internal shocks to occur
(the actual number depends on the amount of variations in the bulk Lorentz factor along the flow).
The minimal value of $g$ (the most efficient) arises when all the initial energy is transformed into two channels:
internal energy of the electrons and energy in magnetic fields.
In this case $g=1/(1-\eta_{int})$. However, if some of the energy remains in thermal energy of the the protons, $g$
would be somewhat larger.

\begin{figure}[h]
\epsscale{0.3}
\centerline{\plotone{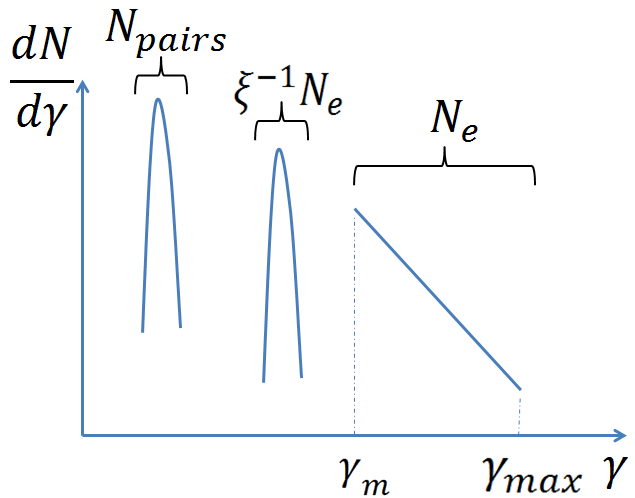}}\caption
{\small The electron distribution is composed of three ingredients: the power-law distributed electrons,
emitting synchrotron at the sub-MeV peak, the thermal ``passive'' electrons which are not accelerated to high Lorentz factors
and the pairs created by annihilation of high energy photons (these too, are not energetic enough to contribute to the sub-MeV spectrum).
Depending on $\xi$, either the second or the third components dominate the total number.}\label{fig:numbers} 
\end{figure}

\begin{figure}[h]
\epsscale{0.5}
\centerline{\plotone{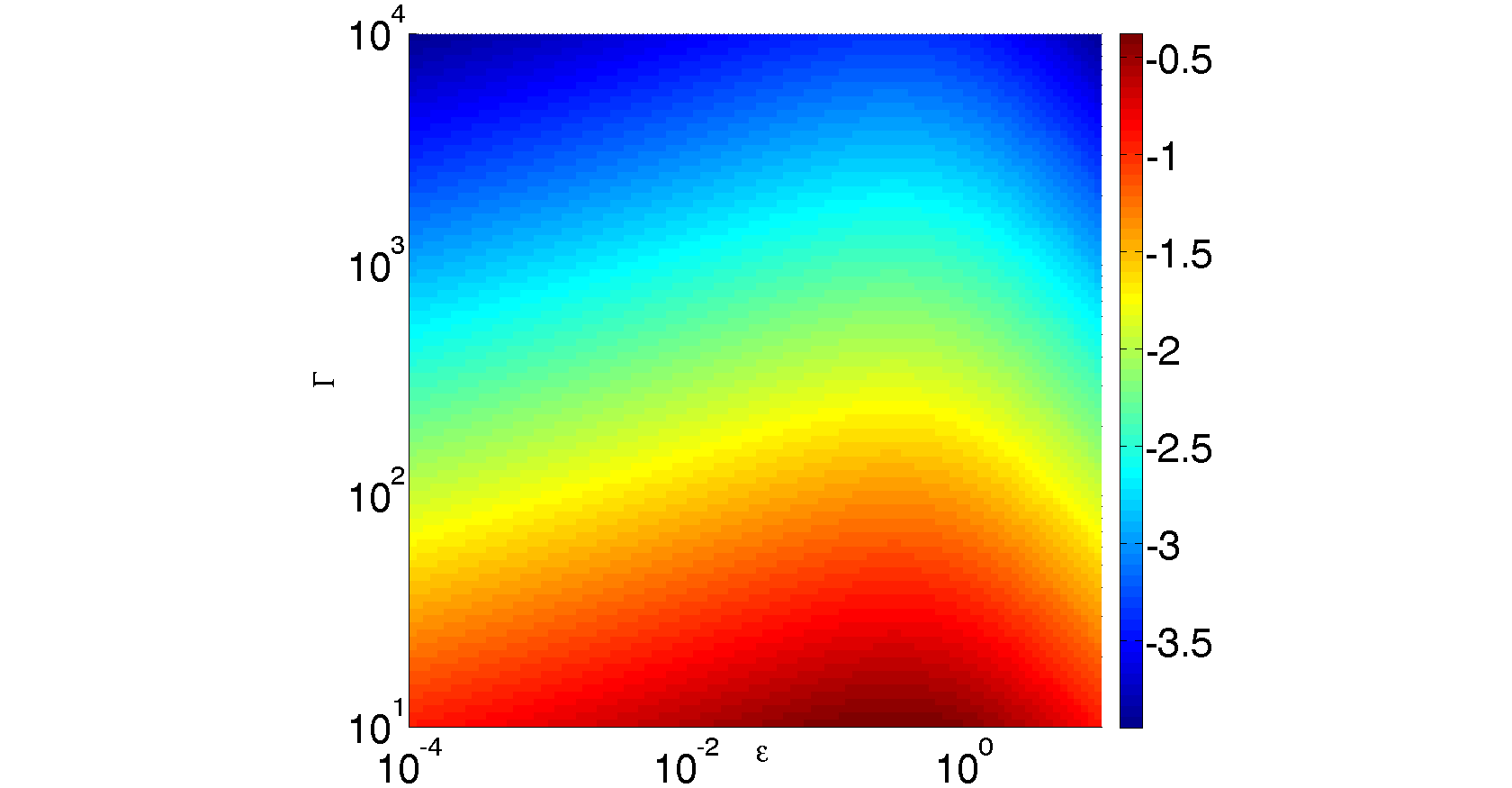}}
{\small $Log[\xi(\G,\epsilon_B,\epsilon_e)]$ of a proton dominated flow within the internal shocks scenario
The results are plotted for $g=2$, $\eta_{int}=0.2$, $p=2.5$ and $\epsilon_B+\epsilon_e=1$. 
This is an upper limit on the actual value of $Log(\xi)$ for real values of $\epsilon_B,\epsilon_e$. Even for modest values of $\G$,
$\xi \ll 1$ which corresponds to a small fraction of relativistic electrons in the emitting region.} \label{fig:number_r}
\end{figure}

\begin{figure}[h]
\epsscale{0.5}
\centerline{\plotone{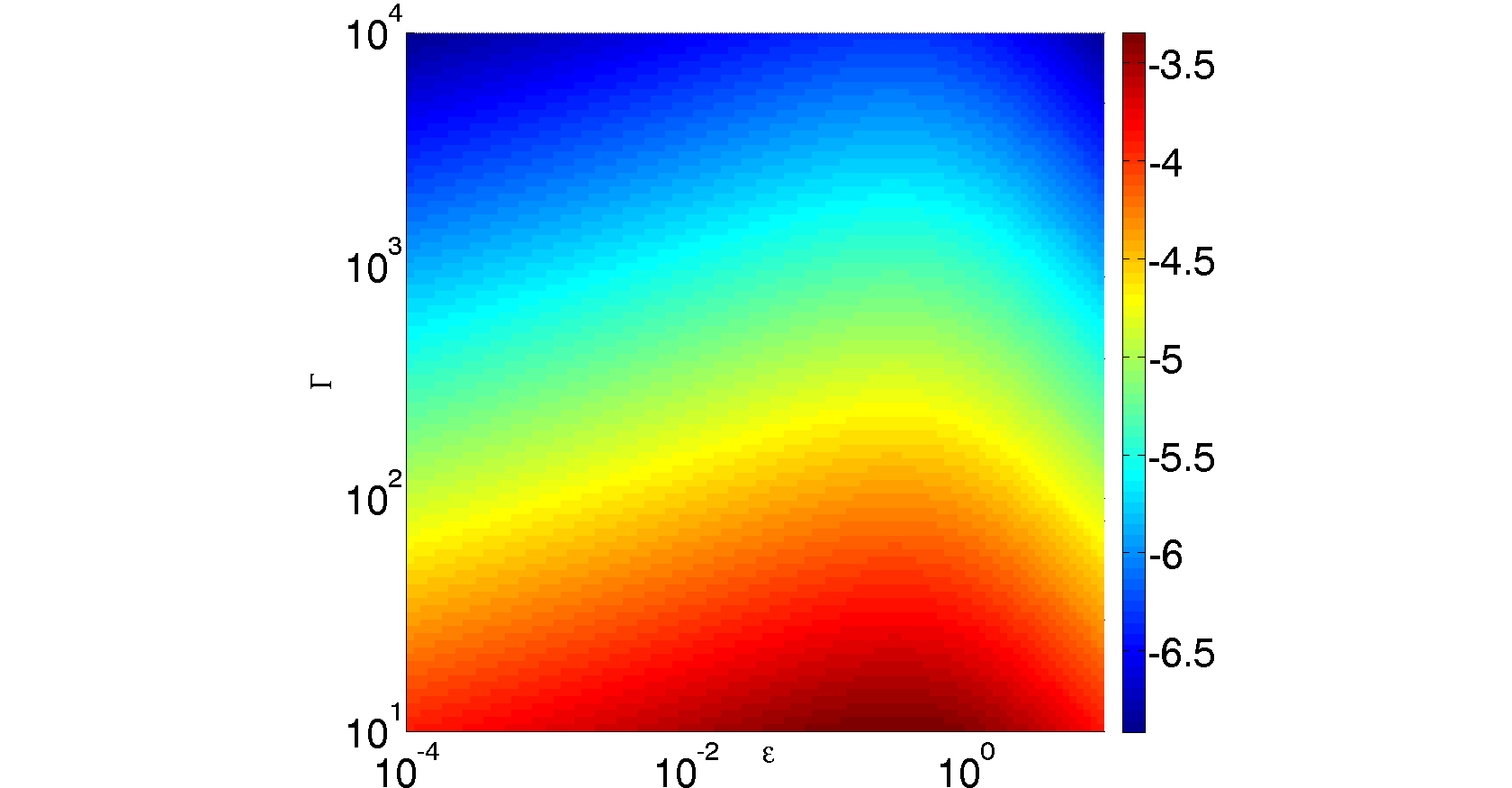}} \caption
{\small Same as Fig.\ref{fig:number_r} but for a pair dominated flow, i.e.
$Log[\xi[\G,\epsilon_B,\epsilon_e)]$ for $g=2$, $\eta_{int}=0.2$, $p=2.5$ and $\epsilon_B+\epsilon_e=1$.
For reasonable values of $\G$,
$\xi \lesssim 2\times10^{-5}$.} \label{fig:number_r2}
\end{figure}

Eq. \ref{Etot} yields the energy of the emitting electrons: $E_e=\epsilon_e E_{int}=\epsilon_{e} \eta_{int} E_{tot}$
in terms of the initial energy stored in the protons (or pairs).
At the same time the synchrotron conditions determine the number of emitting electrons, $N_e$ and
their Lorentz factor, $\gamma_m$ and this determines, in turn, the energy of the emitting electrons via Eq. \ref{electronsE}.
Comparing the two expressions we find that the number of emitting electrons is significantly
lower than the total number of electrons. We define $\xi$ as the fraction of the synchrotron emitting electrons
in the flow: $\xi \equiv N_e/N_{tot}$. A schematic picture of the electron number distribution is given in Fig. \ref{fig:numbers}.

\begin{equation}
\label{E_e}
E_e=N_e \frac{p-1}{p-2} \G \gamma_m m_e c^2=\xi N_{tot}\frac{p-1}{p-2} \G \gamma_m m_e c^2
\end{equation}
Using Eqns. \ref{N_{rel}}, \ref{gamma}, \ref{Etot}, \ref{E_e}, $E_e=\epsilon_e E_{int}=\epsilon_{e} \eta_{int} E_{tot}$ becomes:
\begin{equation}
\label{numbers}
\xi(\G,\epsilon_B,\epsilon_e)=\frac{N_e}{N_{tot}}=0.19\frac{p-2}{p-1}\frac{g}{2} \frac{\eta_{int}}{0.2} (1+Y_0)^{1/4}\G_2^{-1} \epsilon^{\frac{1}{4}}\epsilon_e k^{-\frac{1}{4}} (1\!+\!k)^{\frac{3}{4}}
\nu_{p,500}^{-\frac{1}{4}} t_{p,.3}^{-\frac{1}{2}}F_{-26.2}^{\frac{1}{4}}(\times \frac{m_e}{m_p} \mbox{for pairs}).
\end{equation}
The reason for these low fractions of emitting electrons is clear.
The synchrotron solution requires relatively large electron
Lorentz factors, of order $3\times 10^3<\gamma_m<10^5$. On the other hand if the available energy $\epsilon_{e} \eta_{int} E_{tot}$
is distributed to all the electrons (whose number is dictated by the initial number of
protons) than the typical Lorentz factor would be of order $g \epsilon_e \eta_{int} (p-2)/(p-1)(m_p/m_e)$
(or $g \epsilon_e \eta_{int} (p-2)/(p-1)$ for a pair-dominated flow).
Unless g is unreasonably large, this Lorentz factors is too small by several orders of magnitude. To overcome this problem we
must conclude, following \cite{Daigne(1998),Bosnjak(2009)} that within the internal shocks model only a small
fraction of the electrons are accelerated in the shocks.
It should be noted, that the conditions needed for this solution, are very different than the situation achieved in recent
PIC simulations. This is because, in order for the solution presented here to work there must be a gap
between the Lorentz factor of the thermal (i.e. non synchrotron emitting) electrons and $\gamma_m$ as depicted in Fig. \ref{fig:numbers}.
The ratio of the number of emitting electrons to the total number is plotted in Fig. \ref{fig:number_r} (Fig. \ref{fig:number_r2} for pairs),
for the limiting case of $\epsilon_B+\epsilon_e=1$
which provides an upper limit on the actual value of $\xi(\G,\epsilon_B,\epsilon_e)$. 
Large values of $\G$ lead to $\xi(\G,\epsilon_B,\epsilon_e)\ll1$.
For $\G>100$ (as found in \S \ref{synchsol}), $\xi \lesssim 4 \times 10^{-2}$ ($\approx2\times10^{-5}$ for pairs).

In case $\xi$ is very small, there are many ``passive'' electrons in the flow which may dominate the optical depth.
$\xi_*$ is defined such that for $\xi<\xi_*$ the passive electrons dominate the total number of electrons:
\begin{equation}
\label{xidivide}
 \xi_*=1.4\times 10^{-3} (1+Y_0)^{5/9}(\epsilon)^{3/8} k^{1/8}(1\!+\!k)^{-3/8}\nu_{p,300}^{5/8} t_{p,.3}^{1/4} F_{-26.2}^{-1/8} d_{28}^{-1/4}.
\end{equation}
In this regime:
\begin{equation}
\tau=1.1\times10^{-2}(1+Y_0)^{5/24}\G_2^{-6}(\epsilon)^{1/4}k^{-1/4}(1\!+\!k)^{11/4} \xi^{-1} \nu_{p,300}^{3/4} t_{p,.3}^{-3/2} F_{-26.2}^{5/4} d_{28}^{5/2}.
\end{equation}

\section{pair creation opacity limit (photons above $\nu_{max}$)}
\label{sec:paircreation}
The limit $\tau_{\gamma,\gamma}(\nu_{max})\leqslant1$ leads to a lower limit on $\G$ \citep{Woods(1995),Piran(1999),Lithwick(2001)}.
Eq. \ref{tau2} shows that in the typical case, most of the electrons in the flow are not emitting the synchrotron signal,
but are produced by pair creation.
Therefore, the limit on $\G$ from the requirement $\tau_{\gamma,\gamma}(\nu_{max})\leqslant1$ is always more constraining than that given by Eq. \ref{tau2}.
In order to find $\tau_{\gamma,\gamma}(\nu_{max})$ we calculate $n_{>max,an}$, the number of photons in the source with enough energy to annihilate
the highest frequency photon, $\nu_{max}$ by integrating over the spectrum.
This yields a lower limit on $\G$:
\begin{equation}
 \G_2>(10^{-(4.6+1.8\beta)} |1+\beta|\nu_{max,1}^{\frac{1}{1+\beta}}(1\!+\!k)^{-2}
 \nu_{p,300}^{\frac{1}{1+\beta}}t_{p,.3}F_{-26.2}^{-1}d_{28}^{-2})^{\frac{2\beta-2}{(1+\beta)^2}}.
\end{equation}
This limit is independent of $\epsilon$.

\begin{figure}
\centering
\epsscale{1.1}
\plottwo{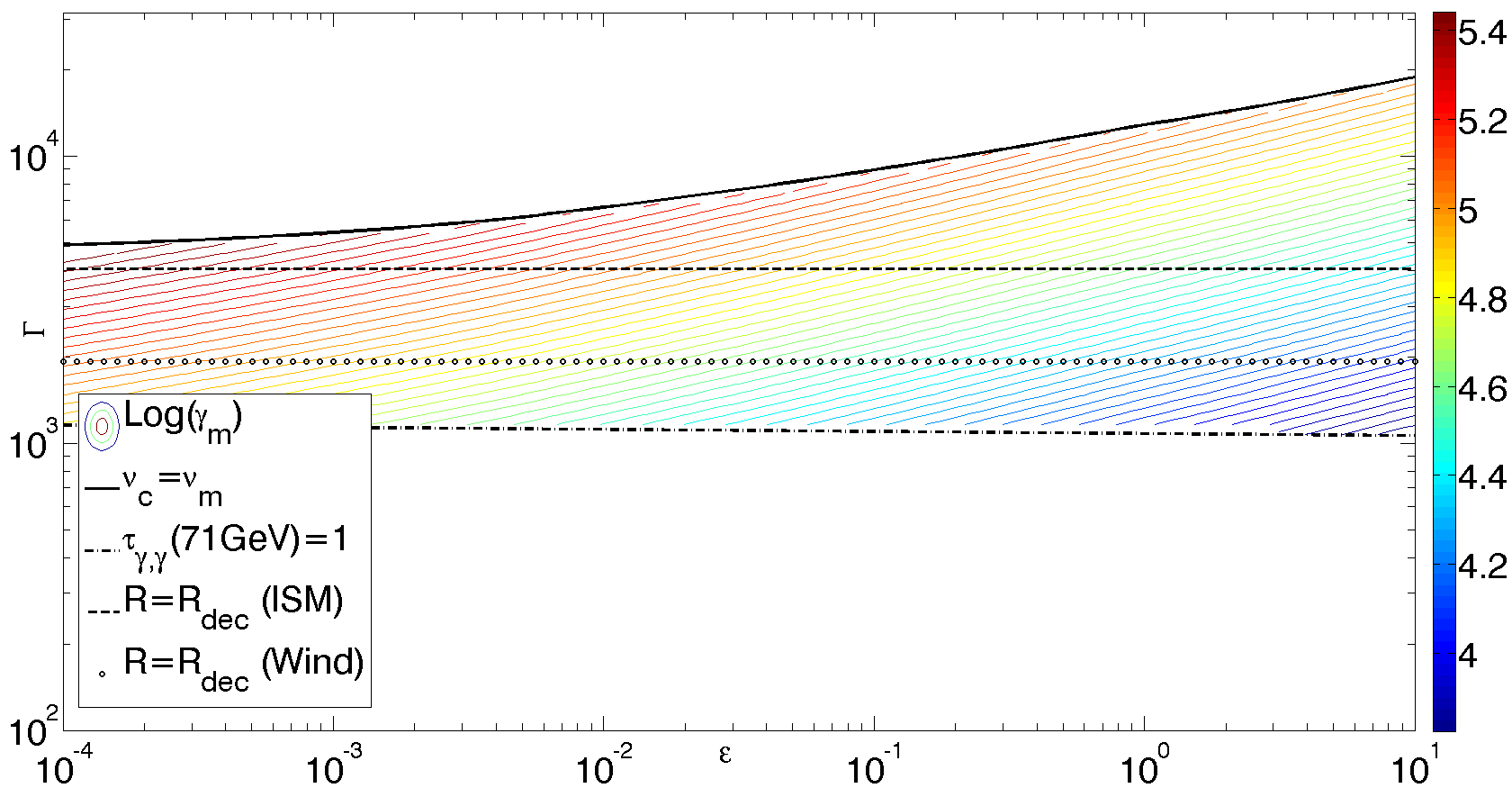}{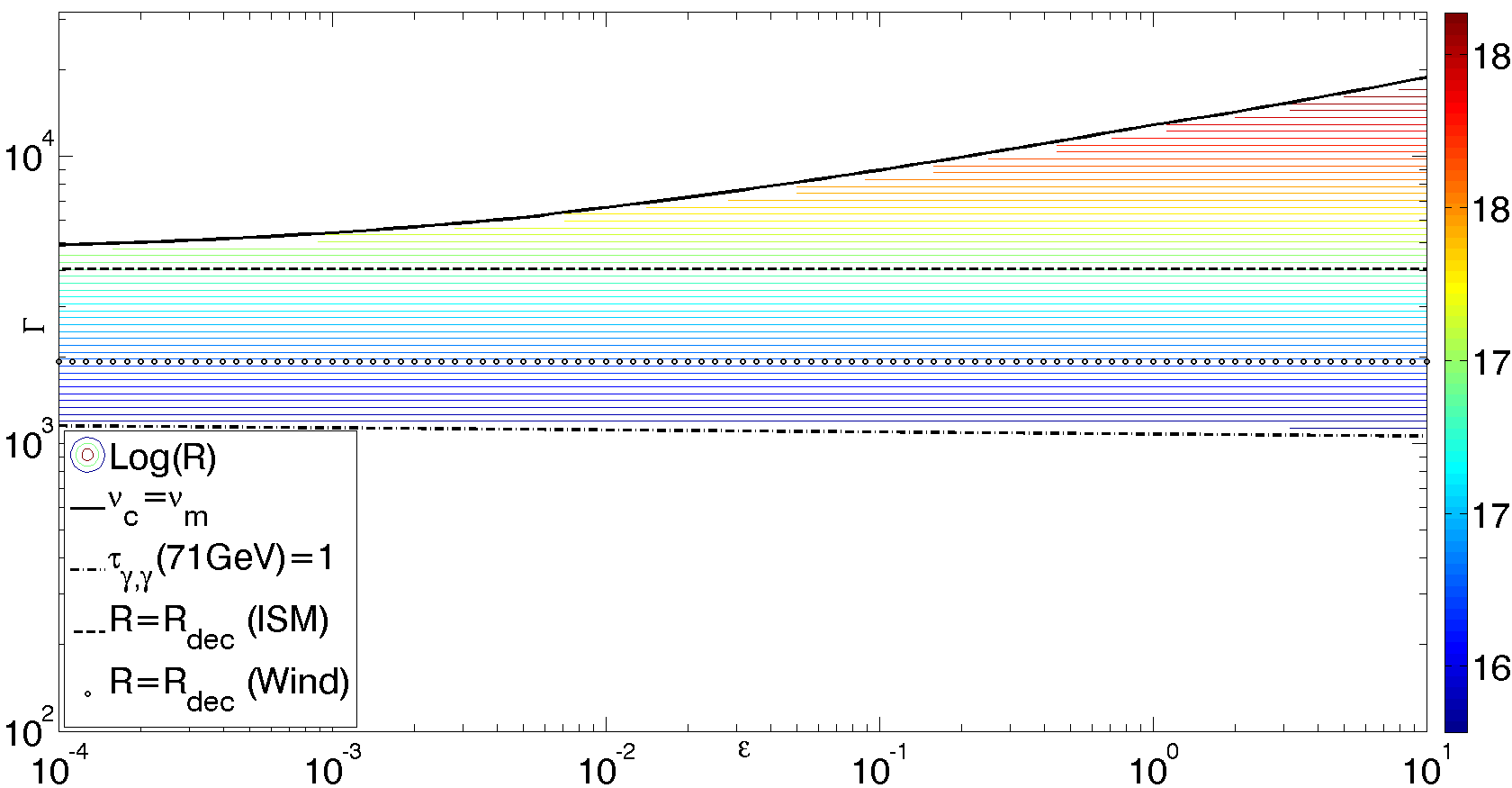}
\caption
{\small The allowed region in the $(\epsilon, \G)$ plane for
GRB 080916C, with $k=1$, $\xi=1$, $h\nu_{max}=71 GeV$ (source frame).
Plotted are: $\gamma_m$ (left panel) and $R$ (right panel).
Colours depict the strength of these parameters.
The conditions $\tau_{\gamma,\gamma}<1$ (bottom horizontal line) and $\nu_c<\nu_m$ (top curved line) impose strict limits
on the parameter space. The area between them is available for Synchrotron solutions.
Two other conditions: $R<R_{dec}(\mbox{Wind})$ and $R<R_{dec}(\mbox{ISM})$ impose softer
limits (the areas below these lines are allowed for each case) on the possible solutions.} \label{fig:ICregion}
\end{figure}

These lower limits on $\G$ replace those used in Figs. \ref{fig:syn_k1}, \ref{fig:syn_k10} in case the high energy photons originate from the same site
as the sub-MeV emission.
$\G\gtrsim300$ is required for typical parameters,
a factor of two larger than the lower limit imposed by considering the optical depth for scattering of synchrotron photons.
Fig. \ref{fig:ICregion} depicts the allowed region in the ($\epsilon,\G$) plane, arising from
the lower limit on $\G$ discussed here and the upper limits given
by Eqns. \ref{coolm}, \ref{decRad}. The results are plotted for the most constraining GRB to date, GRB 080916C,
for which $h\nu_{max}=71 GeV$ (source frame) was observed.
For this burst, we find very high values of the Lorentz factors: $10^3<\G<10^4$, $10^4<\gamma_m<3 \times 10^5$,
and a large emitting radius $4 \times 10^{16}\mbox{cm}<R<3 \times 10^{18}\mbox{cm}$ that is only marginally compatible
with the canonical wind parameter.
Even for this extreme case, the synchrotron solution is possible.
In fact, the allowed parameter space may be even larger in case the high energy emission originates from a different zone than
the sub-MeV radiation \citep{Zou(2009), Zou(2011), Hascoet(2012)}.

\begin{figure}
\centering
\plottwo{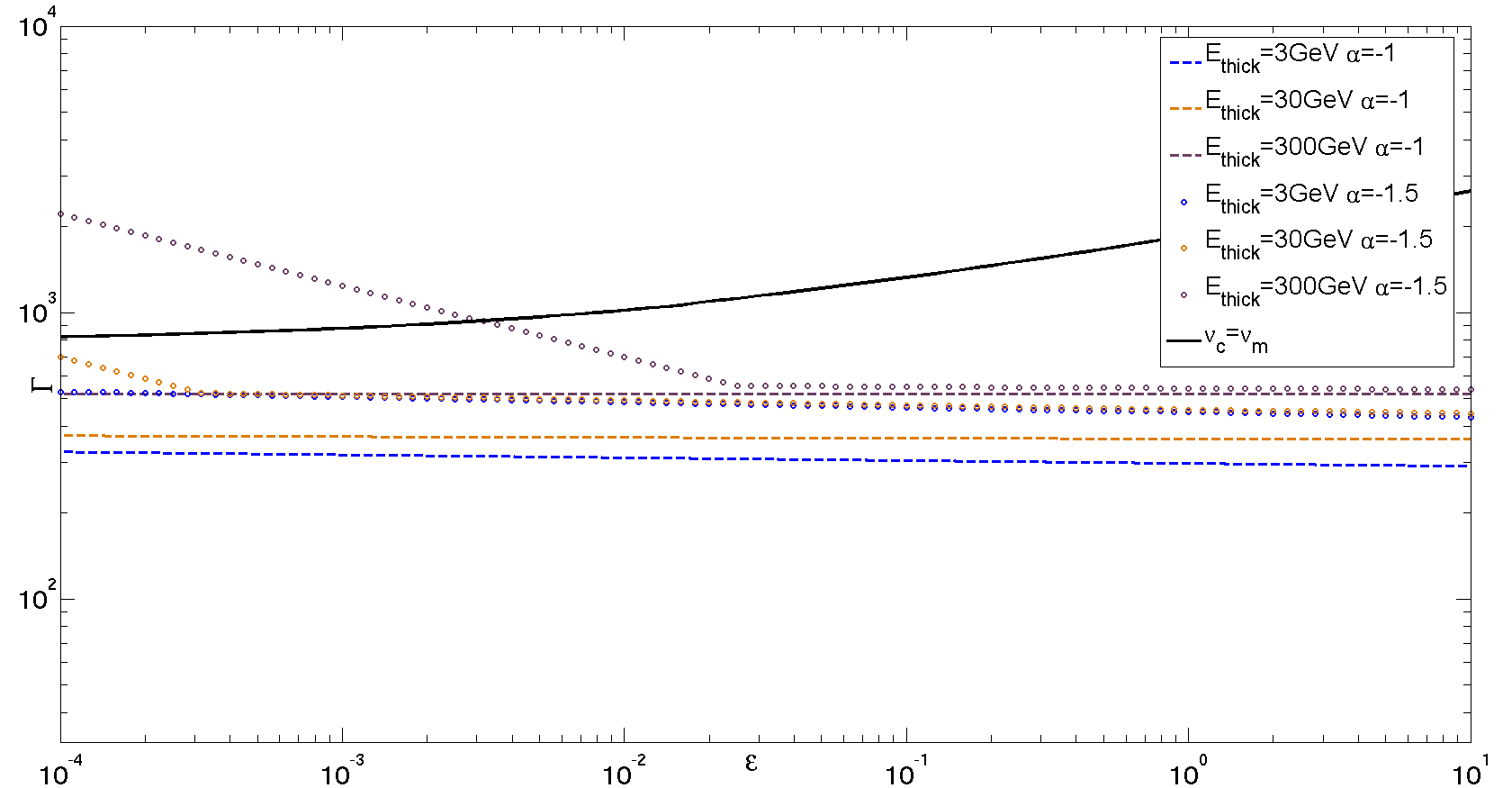}{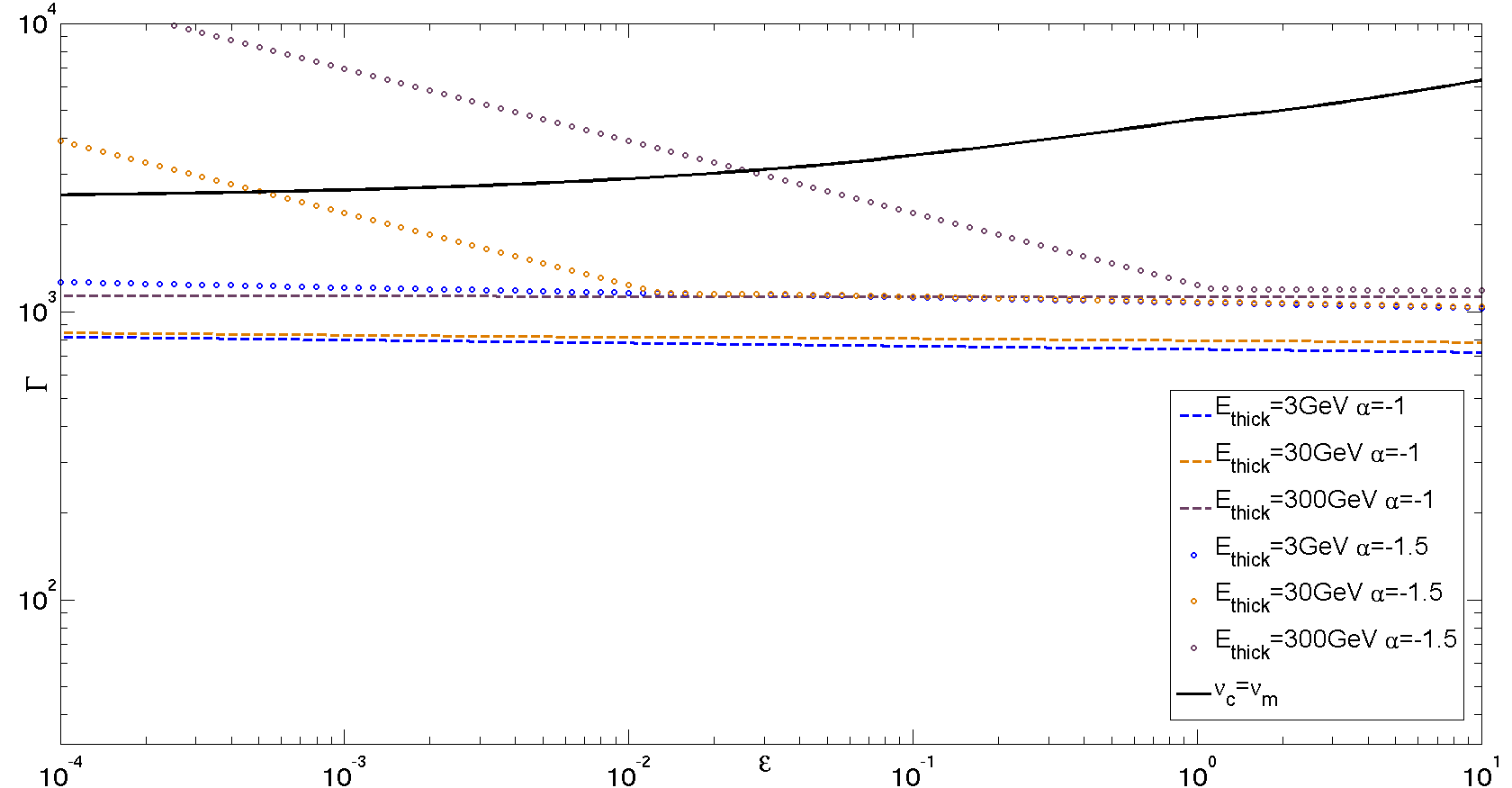}
\caption
{\small Lower limits on $\G$ from the combined constraints on up-scattered flux and pair creation opacity shwon along with
the upper limit on $\G$ originating from $\nu_c=\nu_m$ for various values of the lower spectral slope ($\alpha$) and the photon
energy at which the system becomes optically thick to pair creation ($E_{thick}$).
The area above between the dashed or dotted lines and the solid line is allowed by observations.
Left panel: $k=1$, right panel: $k=10$.} \label{fig:Glimit1}
\end{figure}

As mentioned earlier, given that no opacity break is seen in the spectrum up to $\nu_{max}$,
the frequency in which the system becomes optically thick to pair creation, $\nu_{thick}$ obeys: $\nu_{thick}>\nu_{max}$.
So far, we still have a large uncertainty in the actual value of $\nu_{thick}$. In what follows, we argue that this uncertainty
can be reduced.
The higher $\nu_{thick}$ lies within the range [$\nu_{max}$,300GeV] (300GeV is the upper limit of the LAT detector),
the implications of the low LAT detection rate becomes more constraining on the SSC peak (it should either peak at higher frequencies
or be weak enough to account for observations).
We therefore plot the combination of this pair creation opacity limit and the limit from the SSC flux in the GeV (discussed in \S \ref{sec:SSC}),
in Fig. \ref{fig:Glimit1} for $k=1$ and $k=10$.
These lower limits on $\G$ are more constraining for lower values of $\alpha$ (causing a smaller loss by extrapolation of the SSC from its
peak to the GeV) or for low values of $\epsilon$ for which the SSC flux is larger.
Quantitatively, this is described by Eq. \ref{Glimit}.
We see that in case $\nu_{thick}$ lies at the highest frequency observed by LAT (or above) and $\alpha=-1.5$,
the SSC limits rule out $\epsilon\lesssim5\times 10^{-3}$ for $k=1$ ($\epsilon\lesssim5\times 10^{-2}$ for $k=10$).

\section{Implications of CTA observations on the Synchrotron model}
\label{sec:CTA}

\begin{figure} [h]
\epsscale{0.4}
\plotone{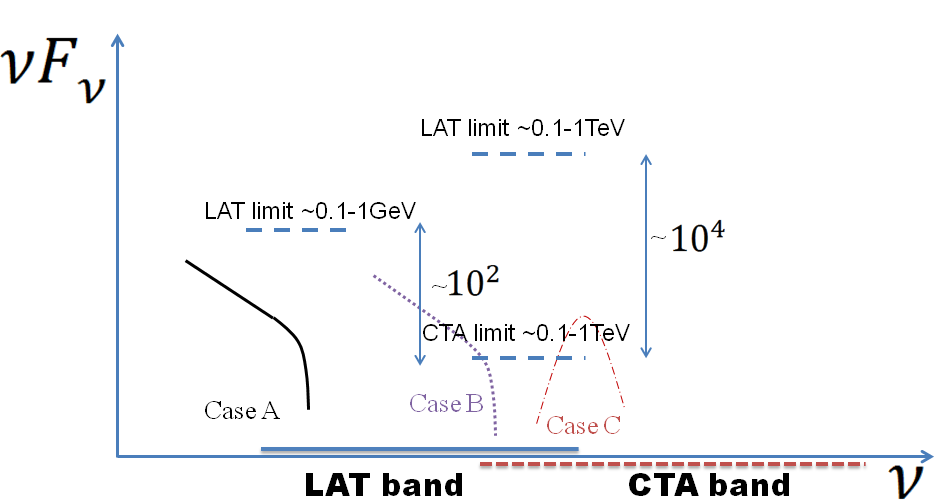} \caption
{\small The three possibilities for CTA observations, described in the text. Case A: No signal in the CTA - no significant costraints on
the parameter space (as this could be due to optical depth). 
Case B: a weak signal at tens of GeV, where the CTA is much more sensitive than LAT, leading to strong constraints on SSC.
Case C: a detection of a peak (in $\nu F_{\nu}$) in the CTA band, leading to a full solution of the SSC model or ruling it out.} \label{fig:CTA}
\end{figure}

It is interesting to explore the implications of future GRB observations by the CTA, in the tens of GeV to tens of TeV energy range,
on the synchrotron model.
The universe is optically thick to radiation above a few hundred GeV \citep{Nikishov(1962), Gould(1966)} and it will therefore be difficult to observe GRBs from typical redshifts
$z\approx1$ in this energy range.  
However, the CTA is expected to have better detectability in the tens of GeV range than LAT by at least four orders of magnitude \citep{CTA(2010), CTA(2011)}.
There are three possible scenarios regarding observations of the prompt emission by CTA, schematically drawn in Fig. \ref{fig:CTA}.
Least intriguing is a null detection.
This could happen either because the system is optically thick to intrinsic pair creation at lower frequencies than the CTA band (i.e. the breaks
in the LAT spectrum are due to optical depth for pair creation, and the fact that we do not see anything with CTA is trivially expected)
or because the breaks in the LAT spectrum are intrinsic spectral breaks due to
the maximal synchrotron frequency and there is simply a very low SSC signal at the CTA band.
The latter case could provide significant constraints on the parameters of the model.
However, since we cannot rule out a large optical depth, we won't obtain new constraints on the model just
from a null detection.
Second, there might be a weak signal in the tens of GeV range which is, however,
significant enough to show that the system is optically thin at least up to the CTA band.
In this scenario we may associate the LAT breaks with the maximal synchrotron frequency and may obtain $\G$ according to Eq. \ref{nusynmax}
up to the numerical factor $f$ which is expected to be of order unity.
In addition, the improved sensitivity of CTA in the tens of GeV range (relative to that of LAT), could possibly allow us to rule out a wide range of SSC solutions
(the SSC component should be both weak and peak at higher frequencies than $\nu_{max}$ for the SSC solution to be compatible with such observations).
Effectively, the improved sensitivity would increase the lower limit on $\G$ given by Eq. \ref{Glimit}, by replacing $\nu_{max}$ by the maximal photon energy observed with the CTA
and replacing the fractional flux detected by LAT, $\eta_{LAT}$ with the corresponding fraction  $\eta_{CTA}$
that could be lower by up to two or three orders of magnitude
(depending on the actual flux that would be detected by CTA and on the new $\nu_{max}$).
This would narrow down the allowed parameter space and would produce lower limits on $\G$ and $\epsilon$.
For instance, a detection of even a few photons in the tens of GeV range, could lead to $\epsilon>0.3$ and $\G>500$.
Finally, a strong signal, with a fluence larger than the extrapolated LAT emission (i.e. with approximately $10^{-7}ergs/cm^2$), 
detected by the CTA would  imply a detection of a new component in this energy band. Such a component, if it ascosiated with the same emitting region,
can naturally be explained as the SSC peak
If the detection is strong enough, it would allow us to measure both the peak flux and the peak frequency (see \S \ref{sec:SSC}).
This will allow, in turn, to determine $\nu_{KN}$ and $F_{SSC}$ and eventually to estimate also
$\G$ and $\epsilon$,
and hence all other parameters of the system up to $k$. 
In case there values turn out to be outside the limits on the parameter space ($\nu_m>\nu_c$, $\tau>1$ etc.) described in \S \ref{synchsol}
these measurements could possibly rule out the synchrotron solution.
In addition if we can measure $\nu_{syn,max}$ we may obtain another equation and obtain the full solution of the model.

\section{Conclusions}
We examined the allowed parameter space for the synchrotron emission for the prompt phase of GRBs.
We considered individual pulses as the building blocks and demanded that the system is fast cooling with
the peak ($\nu F_{\nu}$) frequency comparable to $\nu_m$ so that the system can generate the observed flux efficiently.
Our discussion is general and it does not depend on the specific energy generation or particle acceleration mechanism.

We characterize the possible parameter space in terms of two parameters: the ratio of magnetic to electron's energy, $\epsilon$ and the bulk Lorentz factor, $\G$.
A third auxiliary parameter is $k$, the ratio between the shell crossing time, and the angular timescale.
Efficiency considerations alone limit the ratio of $\epsilon$: $10^{-4}<\epsilon<10$
(see \S \ref{energy_efficiency}). The upper limit arises from the requirement that a large fraction of the total kinetic energy has to  
be stored at some stage in the relativistic electrons ($\epsilon_e\gtrsim0.1$) and thus be available for radiation.
The lower limit arises from the requirement that the up-scattered inverse Compton radiation does not dominate over the energy of the observed sub-MeV pulse.
The lower limit becomes more stringent for shorter pulses and softer pulses
(for example, $t_p\approx5\times 10^{-2}$sec, $\nu_p\approx100$KeV lead to $8 \times 10^{-3}<\epsilon<10$).
Another implication of efficiency requirements is that soft - high luminosity bursts require huge intrinsic powers in the synchrotron model.
This may explain why such bursts are not observed. 

Efficiency also constrains $\G$.
The upper limit is determined by the condition that the emission process is efficient, namely that 
$\nu_c\lesssim\nu_m$. This is more constraining for small values of $\epsilon$ and 
therefore it suggests rather strong magnetization.
The lower limit on $\G$ arises from limits on the SSC component and from opacity considerations.
These lead to $150<\G<3000$. Both the limits increase slightly with k, namely when the emitting radius
is larger compared with the radius implied by variability.
It is interesting to observe (see e.g. Figs. \ref{fig:syn_k1},\ref{fig:syn_k10}) that within the allowed solution 
$\G$ is not necessarily correlated with  $\epsilon$).

Current Observations of high energy (GeV) photons lead to  pair opacity constraints and to a larger lower limit on $\Gamma$. 
However, this limit depends on the fact that the high frequency photons originate at the same site as the sub-MeV photons (or at smaller radii)
and it maybe relaxed if this is not the case.
The observations of high energy photons actually only produces lower limits on the true frequency at which the system becomes
optically thick for pair creation ($\nu_{thick}$). We find that if, in reality, this frequency is as high as 300GeV,
then low $\alpha$ bursts push the model further towards equipartition ($\epsilon>5\times 10^{-3}$ for $k=1$
or $\epsilon>5\times 10^{-2}$ for $k=10$).

We explore the implications on the synchrotron model of future high energy observations by the CTA,
in the tens of GeV to tens of TeV energy range.
We find three general possibilities.
Most trivially, a null detection by CTA does not yield significant constraints on the model.
Second, a weak detection would imply that $\nu_{thick}$ is at least of the
order of tens of GeV, and that there is a weak SSC signal up to this frequency.
This increases the lower limits on $\G$ and $\epsilon$.
Depending on the actual flux measured, these limits can be as strong as $\epsilon>0.3$ and $\G>500$.
Third, a strong signal in the CTA band (above the extrapolation of the LAT signal) would be suggestive of an SSC peak.
This could allow us to determine at least two of the three free parameters of our model ($\Gamma$, $\epsilon$, $k$)
and effectively determine the full solution or possibly
rule out the synchrotron model in case either of the first two parameters turns out to be beyond the limits discussed in this paper.

Even though we don't have a full solution we can obtain
a reasonable estimate of the expected range of synchrotron parameters needed to produce the observed prompt emission.
The emission radius, $R$, is  relatively large:
$10^{15}\mbox{cm}<R<10^{17}\mbox{cm}$ and it is independent of $\epsilon$. 
The magnetic field,  
$B$, spans almost four orders of magnitude: $3\mbox{ Gauss}<B<10^4\mbox{ Gauss}$.
The lines of constant $B$ are almost parallel to the $\nu_m=\nu_c$ line. Larger values of B lead to faster cooling so that the value of $B$ is
almost proportional to $x_{ins}$, the ratio between the number of
instantaneously emitting relativistic electrons and the overall number of relativistic electrons.
The allowed range for $\gamma_m$ is: $3\times 10^3<\gamma_m<10^5$. These large values of $\gamma_m$ (compared with $m_p/m_e$) imply  that, in many cases, as we discuss below,
only a small fraction of the electrons ($10^{50}<N_e<10^{52}$)
are accelerated to relativistic velocities and participate in the synchrotron emission process.
These large values of $\gamma_m$ also suggest that any SSC peak 
would be at energies of at least a few hundred GeV. This is  above the LAT band, 
naturally causing the SSC signal to be hard to observe with current detectors.
Interestingly these parameter ranges, are very weakly dependent on $k$
and they remain within this range as long as we keep the observed quantities at their canonical values.

Slightly deviating from the generic approach we present here, we examined the implications of the model 
if most of the energy is stored at some stage in the form of kinetic bulk motion.
This would be naturally the case in the internal shocks model but also if the jet is initially Poynting flux dominated but 
it converts it energy to kinetic energy before the energy is radiated away.
In this case the synchrotron solution requires that
only a small fraction of the electrons in the flow are heated by the shocks and
dissipate their energy during the prompt emission \citep{Daigne(1998),Bosnjak(2009)}.
However, we note that the distribution of electrons needed for this type of solution
is different than what was achieved by recent PIC simulations, as it requires a gap between the energy of the non-accelerated electrons,
and the minimal energy of the power-law component (see \ref{fig:numbers}).
We find that the fraction of accelerated electrons, $\xi$ must satisfy $\xi\lesssim4\times10^{-2}k^{1/2}$ for a proton dominated
flow (see Fig. \ref{fig:number_r}) and $\xi\lesssim2\times10^{-5}k^{1/2}$ (Fig. \ref{fig:number_r2})
for a pair dominated flow.

The main caveat of this model, that we did not address here, remains the ``line of death`` problem, regarding the low energy spectral slope
(see also \citealt{Daigne(2011)}).
A partial solution to this problem can be achieved if the system emits in a marginally fast cooling regime, where $\nu_m\approx\nu_c$.
The high $\G$ values required by opacity considerations, push the model towards this regime.
However, even the marginally fast solution, only steepens $\alpha$ up to $-2/3$ and it does not solve the entire problem.
A different option is to have a steepening of the slope due to high self absorption frequency \citep{Granot(2000)}.
This is not easily achieved, as the self absorption frequency is typically four orders of magnitude below the peak.

Overall we conclude that there is reasonable range in the parameter ($\G$ and $\epsilon$) phase space 
for which a synchrotron solution that fits the peak frequency and the peak flux is possible. 
The main characteristics of this solution are large emission radii, relatively high magnetization and
large values of electrons' Lorentz factors. These Lorentz factors naturally explain the lack of a SSC signal
within the LAT band. In addition, the solution naturally explains the lack of soft - high luminosity bursts.
Possible future detections of a SSC signal at higher frequencies,
will allow to either obtain the full solution of the model, or rule it out altogether.

We thank Jonathan Granot, Rodolfo Barniol Duran and Indrek Vurm for many helpful discussions.
The research was supported by an ERC advanced research grant (GRB). 

\end{document}